\documentclass[sigconf]{acmart}

\AtBeginDocument{%
  }

\setcopyright{acmlicensed}
\copyrightyear{2018}
\acmYear{2018}
\acmDOI{XXXXXXX.XXXXXXX}

\acmConference[Conference acronym 'XX]{Make sure to enter the correct
  conference title from your rights confirmation email}{June 03--05,
  2018}{Woodstock, NY}
\acmISBN{978-1-4503-XXXX-X/18/06}

\usepackage{booktabs} 

\usepackage{graphicx}
\usepackage{subfiles}
\usepackage{array}
\usepackage{amsmath}
\usepackage[ruled,linesnumbered]{algorithm2e}
\usepackage{hyperref}
\usepackage{multirow}
\usepackage{subfigure}
\usepackage{caption}
\usepackage{flushend}
\usepackage{balance}
\usepackage{diagbox}
\usepackage{colortbl}
 \usepackage{enumitem}
 \usepackage{balance}
 \usepackage{makecell}
 \usepackage{float}
\usepackage{fancyhdr}
\usepackage{threeparttable}

\usepackage{url}
\usepackage{bm}
\usepackage{comment}
\usepackage{graphicx}
\usepackage{pifont}
\usepackage{tcolorbox}
\usepackage{subfigure}

\usepackage{listings}

\newcommand{\tool}{\textsc{PatUntrack}}

\newcommand{\revise}[1]{{#1}}

\begin{document}

\title{{\tool}: Automated Generating Patch Examples for Issue Reports without Tracked Insecure Code}

\author{Ziyou Jiang$^{1,2,3}$, Lin Shi$^{4}$, Guowei Yang$^{5}$, Qing Wang$^{1,2,3}$}

\affiliation{$^{1}$State Key Laboratory of Intelligent Game, Beijing, China;\\
$^{2}$Science and Technology on Integrated Information System Laboratory, \\
Institute of Software Chinese Academy of Sciences, Beijing, China;\\
$^{3}$University of Chinese Academy of Sciences, Beijing, China;\\
$^{4}$School of Software, Beihang University, Beijing, China;\\
$^{5}$The University of Queensland, Brisbane, Australia \\
\{ziyou2019, wq\}@iscas.ac.cn, shilin@buaa.edu.cn, guowei.yang@uq.edu.au\\
\country{}}

\authornote{Corresponding author.\\ }

\renewcommand{\shortauthors}{Shi et al.}

\renewcommand{\shortauthors}{Jiang et al.}


\begin{abstract}
{Security patches are essential for enhancing the stability and robustness of projects in the open-source software community.
While vulnerabilities are officially expected to be patched before being disclosed, patching vulnerabilities is complicated and remains a struggle for many organizations.
To patch vulnerabilities, security practitioners typically track vulnerable issue reports (IRs), and analyze their relevant insecure code to generate potential patches.
However, the relevant insecure code may not be explicitly specified and practitioners cannot track the insecure code in the repositories, thus limiting their ability to generate patches. In such cases, providing examples of insecure code and
the corresponding patches would benefit the security developers to better locate and resolve the actual insecure code. 
In this paper, we propose {\tool}, an automated approach to generating patch examples from IRs without tracked insecure code.
{\tool} utilizes auto-prompting to optimize the Large Language Model (LLM) to make it applicable for analyzing the vulnerabilities described in IRs and generating appropriate patch examples. Specifically, it first generates the completed description of the Vulnerability-Triggering Path (VTP) from vulnerable IRs. Then, it corrects potential hallucinations in the VTP description with external golden knowledge. 
Finally, it generates Top-$K$ pairs of \textit{Insecure Code and Patch Example} based on the corrected VTP description.
To evaluate the performance of {\tool}, we conducted experiments on 5,465 vulnerable IRs.
The experimental results show that {\tool} can obtain the highest performance and improve the traditional LLM baselines by +17.7\% (MatchFix) and +14.6\% (Fix@10) on average in patch example generation.
Furthermore, {\tool} was applied to generate patch examples for 76 newly disclosed vulnerable IRs. 27 out of 37 replies from the authors of these IRs confirmed the usefulness of
the patch examples generated by \tool, indicating that they
can benefit from these examples for patching the vulnerabilities.
}
\end{abstract}

\maketitle

\section{Introduction}

{Security patches are essential for enhancing the stability and robustness of projects in the Open-Source Software (OSS) community. 
In 2017, the Software Engineering Institute (SEI) at Carnegie Mellon University released the {CERT Guide to Coordinated Vulnerability Disclosure (CVD)}~\cite{cert_guidance}, which officially states that individuals or organizations should "\textit{deploy a patch or take other remediation action}" before they disclose a vulnerability to the public security databases~\cite{iso_disclosure,householder2017cert}, such as Common Vulnerabilities and Exposure (CVE)~\cite{CVE}.
However, patching vulnerabilities is complicated and remains a struggle for many organizations~\cite{DBLP:journals/queue/ODell17}.
For example, a vulnerability in the project \textit{python-markdown2} has "\textit{no fix}" and "\textit{welcome pull requests}" for a long time, as is reported in the issue \textit{trentm/python-markdown2/issues/285}~\cite{example_patch_difficulty}. 
These unpatched vulnerabilities can be utilized by attackers through deploying exploits to harm the affected systems (e.g., zero-day~\cite{DBLP:conf/ccs/BilgeD12} and one-day~\cite{DBLP:conf/noms/ElbazRM20,DBLP:journals/corr/abs-2404-08144} attacks), resulting in millions dollars of business losses~\cite{article}.
}



{OSS developers typically report vulnerabilities through the issue reports (IRs)~\cite{DBLP:conf/esem/WaldenDWW09}. Security practitioners, who manage the vulnerability disclosure, can then track these IRs with issue-tracking systems~\cite{bugzilla,jira}, and analyze the relevant insecure code to generate potential patches.
However, the relevant insecure code may not be explicitly specified and practitioners cannot track the insecure code in the repositories, thus limiting their ability to generate patches.
}
{In such cases, providing examples of insecure code and the corresponding patches would benefit the security developers to better locate and patch the actual insecure code.}
{In general, generating example insecure code and patches is challenging, mainly due to the semantic gaps between code and natural language, as well as the information omission in developer's description of the venerability in the IRs. 
Our preliminary study in Section \ref{sec:preliminary} shows that 69.0\% of vulnerable IRs have no insecure code tracked with either manual analysis or State-of-the-Art (SOTA) commit trackers~\cite{DBLP:conf/kbse/ZhangWWXWLJ23, DBLP:journals/jss/RuanCPZ19}, and over 70\% of them were successfully exploited by attackers.
These results inspire us to design an automated approach to generate patch examples based on the IRs without tracked insecure code, which can help security practitioners patch vulnerabilities soon after the IRs are created by the developers.}

{In this paper, we propose an automatic approach, i.e., {\tool}, which generates patch examples from IRs without tracked insecure code.
It optimizes the Large Language Model (LLM) with auto-prompting~\cite{DBLP:conf/emnlp/ShinRLWS20} to make it applicable for analyzing the types and triggering logic of vulnerabilities from their textual descriptions, and generating appropriate patches. {\tool} consists of three main steps: \ding{182} First, it generates the description of the Vulnerability Triggering Path (VTP) from IR, which captures how the vulnerability is triggered. 
\ding{183} Second, it corrects potential hallucinations in the VTP description with external golden knowledge.
\ding{184} Third, it utilizes the VTP description to predict the patch types and generates Top-$K$ pairs \textit{Insecure Code and Patch Example}. 
}

{To evaluate the performance of {\tool}, we conducted experiments on 5,465 vulnerable IRs. 
The results show that {\tool} achieves the highest performance and improves the traditional LLM baselines by +17.7\% (MatchFix) and +14.6\% (Fix@10) on average in patch example generation.
{Furthermore, we applied {\tool} to generate patch examples for 76 newly disclosed vulnerable IRs that do not have tracked insecure code, and asked the authors whether our patch examples can assist them with patching the vulnerabilities.} 
We have received replies from the authors for 37 IRs, and 27 replies 
{confirmed the usefulness of the patch examples generated by \tool, indicating that they can benefit from these examples for patching the vulnerabilities.}
}
To summarize, this paper makes the following main contributions:
\begin{itemize}[leftmargin=*]
    \item 
    \textbf{Technique}: {\tool}, an automated approach to {generate} patch examples for vulnerable IRs {without} tracked insecure code. To the best of our knowledge, {this is the first work on generating patch examples without guidance from the source code.}
    \item 
    \textbf{Evaluation}: An experimental evaluation of {\tool} that shows that {\tool} outperforms all baselines on generating insecure code \& patch examples, {as well as a human evaluation on newly-disclosed vulnerable IRs that further demonstrates its usefulness in practice.}
    \item \textbf{Data}:
    {The datasets and source code~\cite{model_data}, which are made publicly available to facilitate the replication and {the application of {\tool} in the more extensive contexts.}}
\end{itemize}

\section{Preliminaries}
In this section, we first conduct the preliminary study by analyzing the time cost of raising patches for the vulnerable IRs and the exploited ratio of IRs with/without tracked insecure commits.
Then, we provide an example to illustrate the motivation of {\tool}.

\begin{figure}[t]
\centering
\includegraphics[width=\columnwidth]{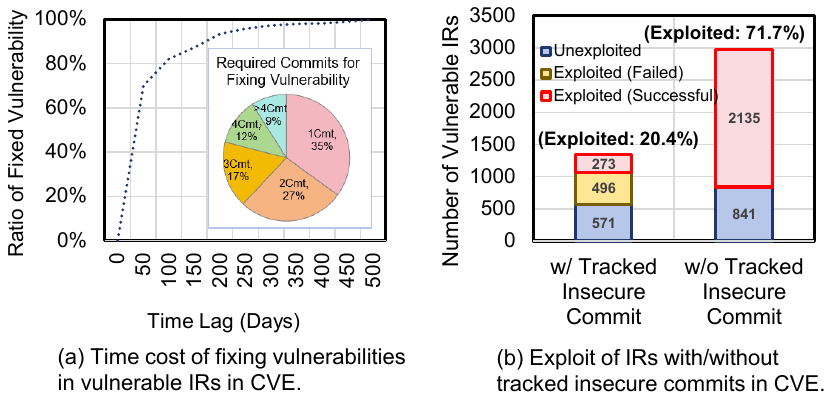}
\vspace{-0.7cm}
\caption{The preliminary study to analyze the time cost of patching and the exploited ratio of the vulnerable IRs.}
\label{fig:preliminary_study}
\vspace{-0.3cm}
\end{figure}


\subsection{Preliminary Study of Vulnerability Patching}\label{sec:preliminary}

To analyze the time cost of vulnerability patching, we introduce a widely-used vulnerability dataset, i.e., GHArchive~\cite{GHArchive}, which 
achieves the IRs in the software community. 
Among them, GHArchive contains vulnerable IRs with their original information of CVE-disclosed vulnerabilities, such as IR's textual descriptions, commits for insecure code and patch (indicated by links with string \texttt{TCommit} and \texttt{TPatch}), etc.
We first analyze the time lag between the IR creation and the vulnerability patching by referring to the commit links to the vulnerable IRs containing insecure commits.
Figure \ref{fig:preliminary_study} (a) shows that nearly 20\% of the vulnerable IRs require over 150 days to raise the patches for successfully fixing the vulnerabilities, and
38\% require over 3 commits to fix the vulnerabilities.

Moreover, we also analyze whether the tracked insecure commits may affect vulnerability exploitation by calculating the intersections of vulnerable IRs with tracked insecure commits and exploited vulnerabilities.
We first decide whether the vulnerable IR contains the insecure commits with the following steps: 
\ding{182} we analyze the commit links in GHArchive;
\ding{183} we utilize the SOTA commit tracker~\cite{DBLP:conf/kbse/ZhangWWXWLJ23} to track the commits if GHArchive does not present the link;
and \ding{184} we manually track the commit links if the previous steps cannot find the commit links.

Second, we determine whether the vulnerability is exploited by analyzing the logs in the links with \texttt{TExploited} for vulnerability exploitation, e.g., exploited time, IP address, and vulnerable version. We find that there are three types of exploitation as follows:
\begin{itemize}[leftmargin=*]
    \item \textbf{Unexploited:} The vulnerabilities are not exploited by attackers.
    \item \textbf{Exploited (Failed):} The vulnerabilities are exploited after they are fixed with specific patches.
    \item \textbf{Exploited (Successful):} The vulnerabilities are exploited before they are fixed, which means the attackers may harm the systems.
\end{itemize}

Figure \ref{fig:preliminary_study} (b) shows that 2,976 of 4,316 vulnerable IRs (69.0\%) cannot track the insecure commits from the GHArchive, 
Among these IRs without tracked insecure code, 71.7\% of vulnerabilities in the vulnerable IRs are successfully exploited by attackers, which is +50.3\% higher than IRs with insecure commits.
These results show that the insecure commits with patches are important to reduce the exploitation of vulnerable IRs, but raising appropriate patches to fix the vulnerabilities is a time-consuming task.





\subsection{Motivating Example}

\begin{figure}[t]
\centering
\includegraphics[width=\columnwidth]{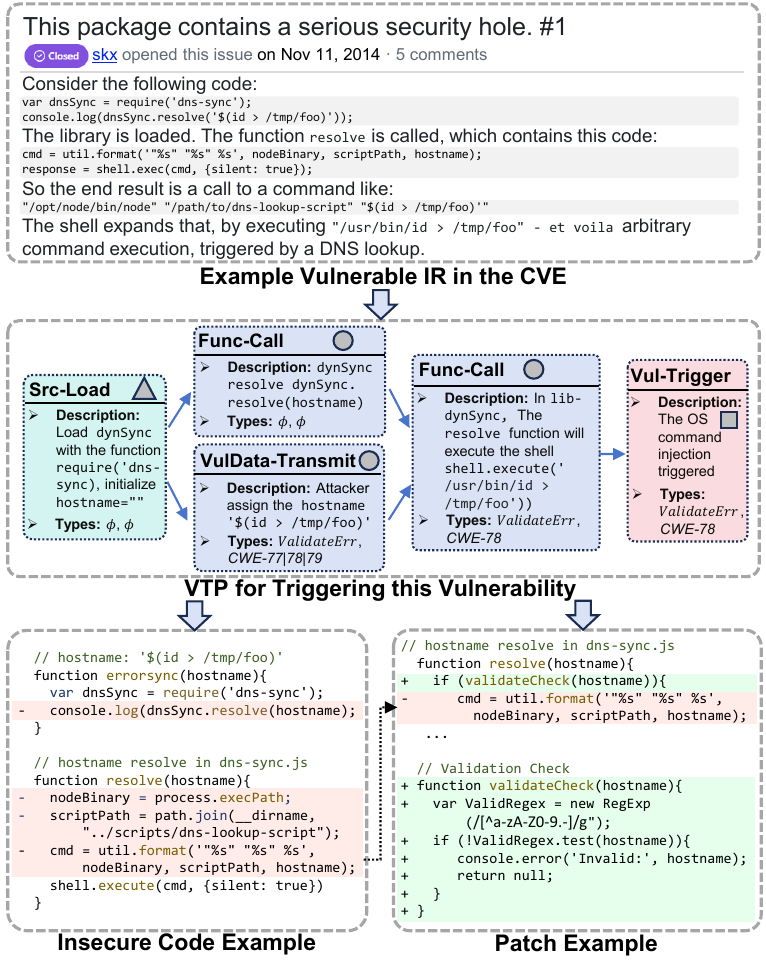}
\vspace{-0.7cm}
\caption{The vulnerable IR and the generated insecure code \& patch example (\textit{skoranga/node-dns-sync/issues/1}~\cite{motivation_example_ir}).}
\label{fig:motivation_example}
\vspace{-0.5cm}
\end{figure}

Figure \ref{fig:motivation_example} shows the motivation of generating insecure code \& patch examples from the vulnerable IR.
From the example, we analyze how the vulnerability is triggered based on the textual description of IR. The project first loads the library \texttt{dynSync} and initializes the parameter \texttt{hostname}. 
Then, it calls the function \texttt{dynSync.resolve}, which will execute the system command to look up the DNS server to resolve the \texttt{hostname}.
However, this function-calling process has vulnerabilities, since the logic of \texttt{resolve} function contains the execution of the system command. If the attackers initialize the \texttt{hostname} with specific strings like "\texttt{\$(id > /tmp/foo)}", it may inject the vulnerabilities and harm the system.

Referring to the secure coding practices in OWASP~\cite{OWASP}, this vulnerability is a typical OS command injection (CWE-78)~\cite{cwe_78} and the type of patching is validating the input with \textit{Regex Testing}, so the patch example will incorporate the validation of the inputs. 
The system commands will not execute if the input strings contain such specific strings, thus preventing the vulnerabilities. 
\section{Approach}

\begin{figure*}[t]
\centering
\includegraphics[width=\textwidth]{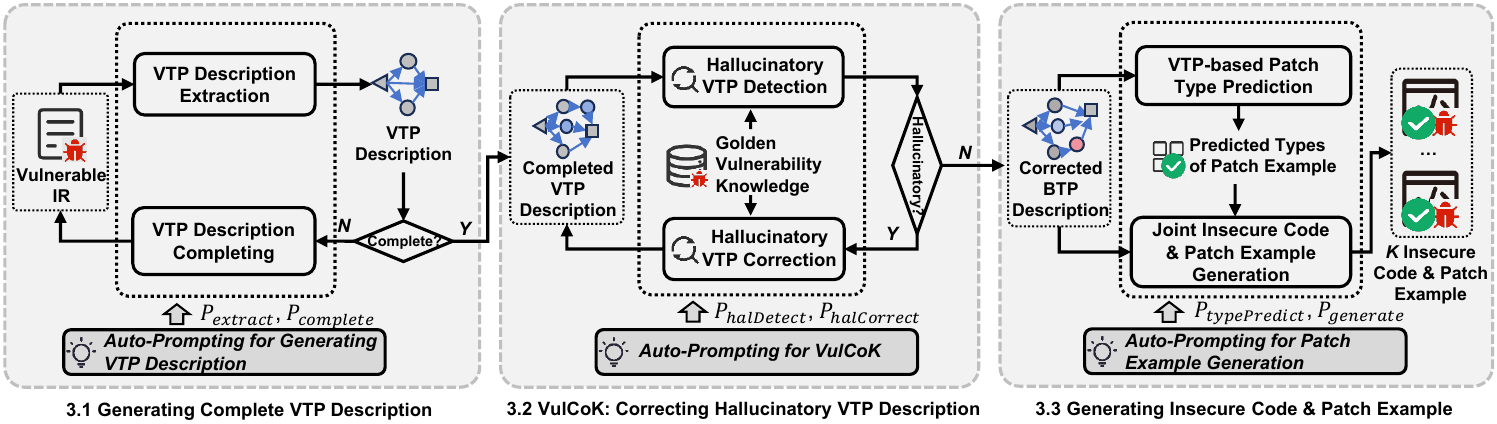}
\vspace{-0.7cm}
\caption{The structure of {\tool}.}
\label{fig:model}
\vspace{-0.4cm}
\end{figure*}

The overall framework of {\tool} is illustrated in Figure \ref{fig:model}.
Since the IR authors may miss some details in describing vulnerabilities, we first extract the VTP descriptions from IR's textual description and complete the missing nodes and edges.
Second, since the pre-trained data and training strategy of LLMs have flaws that result in hallucinations, we propose VulCoK to correct the hallucinations in VTP descriptions.
Third, we jointly generate insecure code \& patch examples with patch type prediction with the corrected VTP description.
For each step, we utilize auto-prompting to optimize LLM to make it applicable to analyze vulnerabilities.

\subsection{Generating Complete VTP Description}\label{sec:vtp_generation}

The IR can be formulated as follows: $\{Title, Body\}$, where $Title$ is the summarization of the main topic;
$Body$ contains a set of sentences that describe the details of IR.
{\tool} generates the complete VTP description by \ding{182} extracting the original VTP description from the IR textual descriptions, and \ding{183} completing the missing nodes and edges to update the vulnerable IRs.

\subsubsection{VTP Description Extraction}

Previously, Cheng et al.'s~\cite{DBLP:journals/tdsc/ChengNLWZS24} define a Bug-Triggering Path (BTP) as a set of program statements to reside in the execution paths toward the location where the error is triggered, which is an effective method to detect and reproduce the vulnerabilities, and has been utilized by different vulnerability detectors~\cite{DBLP:conf/ndss/LiZXO0WDZ18,DBLP:journals/tdsc/0027ZX0ZC22,DBLP:conf/ijcnn/NguyenLVMGP21}.
However, there are gaps between normal bugs and vulnerabilities, so traditional BTPs are not useful to accurately find vulnerabilities.
For example, the traditional BTP defines the operations of the program as \textit{package loading}, \textit{variable declaration}, and \textit{function calling}, which are operations to trigger normal bugs in the OSS projects.
On the contrary, vulnerabilities focus on the transmission of tainted data~\cite{DBLP:conf/icse/ChiangLZBSLNT24}, where we find the "\textit{source}" and "\textit{sink}" code to describe the transmission path of the tainted data and locate the code lines that may produce the vulnerabilities.

The VTP description in our work can identify the triggering paths of vulnerability by incorporating the transmission of tainted data.
The structure can be modeled as $\mathcal{G}_{VTP}=\langle\mathcal{V}_{VTP},\mathcal{E}_{VTP}\rangle$, where the $\mathcal{V}_{VTP}=\{Op_0, Op_1, ..., Op_{T}\}$ is a series of nodes that describe the operations that may result in the vulnerability.
The $Op_0$ is the start operation, which contains the operation to load the sources, such as \textit{loading third-party library} and \textit{initializing variables}, and $Op_T$ is the end operation that indicates the vulnerability is triggered after the previous operations are conducted.

Based on the previous works~\cite{DBLP:journals/tdsc/ChengNLWZS24} and our manual analysis on over 1K vulnerabilities, 
we summarize four types of operations that cover the triggering process of vulnerabilities.
we manually analyzed the IRs with experienced security practitioners who participated in our data annotation to determine the types of operation nodes/edges with \textbf{Open Card Sorting}~\cite{cardsorting1}, a flexible classification method that allows us to create information categories freely, thus helping designers develop more appropriate types.

\begin{itemize}[leftmargin=*]
    \item \textbf{Src-Load:} This operation indicates that the program loads the vulnerability-related source data, such as loading the packages that may have vulnerabilities and loading the variables that may contain the taint information.
    \item \textbf{Func-Call:} Since some vulnerabilities are directly caused by the incorrect calling of the functions, such as the use-after-free~\cite{DBLP:conf/uss/YagemannCSL23} vulnerability, we specifically analyze the function calling processes that may trigger the vulnerabilities.
    \item \textbf{VulData-Transmit:} This operation denotes the transmission of the vulnerable data. The data comes from the source variables or the different libraries, and will finally transmit to the sink code lines that may harm the system.
    \item \textbf{SecData-Transmit:} This operation indicates the transmission of other data in the function calling process. Different from the VulData-Transmit, these data are secure and will not harm the system, and we also analyze the transmission of secure data to distinguish it from vulnerable data.
    \item \textbf{Vul-Trigger:} We add this node to intuitively indicate the results of vulnerability triggering.
\end{itemize}

The edges in VTP description $(Op_i\rightarrow Op_j)\in \mathcal{E}_{VTP}$ is the one-way link that denotes the transition of how to trigger the vulnerability, where $Op_i$ is the prerequisite operation for the $Op_j$.
With the extracted VTP description, we can generate the insecure coding example which can reflect the vulnerable code in the projects, and generate patches based on the coding examples.


Each node $Op_i$ is a triplet $\langle Op\_Type, Op\_Desc, Vul\_Type\rangle$, where
\ding{182} The element $Op\_Type$ is the previous type of operation node. 
\ding{183} The $Op\_Desc$ is the description of the operations, which briefly explains the information of each operation step with texts and a few code snippets.
\ding{184} The $Vul\_Type$ is the type of vulnerability, and we introduce the \textit{CWE type} and the detailed \textit{error types} in it to describe the vulnerability. 
Chow et al.~\cite{DBLP:conf/icse/ChowGP24} indicates that for different error types, the \textbf{focuses} of bug triggering methods are different.
Inspired by it, we define seven error types in VTP based on our manual analysis of vulnerabilities,
as shown in
Table \ref{tab:error_types}. 

\begin{table}[htbp]
\vspace{-0.4cm}
\caption{The error types in the VTB operation nodes.}
\vspace{-0.4cm}
\resizebox{\columnwidth}{!}{
\begin{tabular}{c|m{4cm}|m{6cm}}
\toprule
\textbf{ID} & \multicolumn{1}{c|}{\textbf{Error Types (VTP)}}              & \multicolumn{1}{c}{\textbf{Description}}                                                                                                                                                                            \\
\midrule
1 & Encoding \& Validation Errors        & The error comes from improperly handled, leading to unexpected behavior.                                   \\
\arrayrulecolor{lightgray}\hline
2& {Dependency Errors}                        & The variables, functions, or libraries are unresolved or incorrect.                                                                       \\
\hline
3&{Injection \& Logic Errors}                & The functions are improperly used leading to incorrect execution.                                                                                \\
\hline
4&\makecell[l]{Memory Management \& \\ Concurrency  Errors} & The way how data is handled has errors, leading to exploitable conditions.                   \\
\hline
5&Race \& Configuration Errors             & The error relates to the system conditions, affecting its correct operation.                                  \\
\hline
6&Buffer overflow                   & The error involves improper handling of boundaries and limits in storage, leading to overflows.    \\
\hline
7&{Error handling \& logging issues}  & The encoding or decoding have errors, leading to XSS or injection attacks.                      \\
\arrayrulecolor{black}\bottomrule
\end{tabular}}
\vspace{-0.2cm}
\label{tab:error_types}
\end{table}

\begin{figure}[htbp]
\centering
\includegraphics[width=\columnwidth]{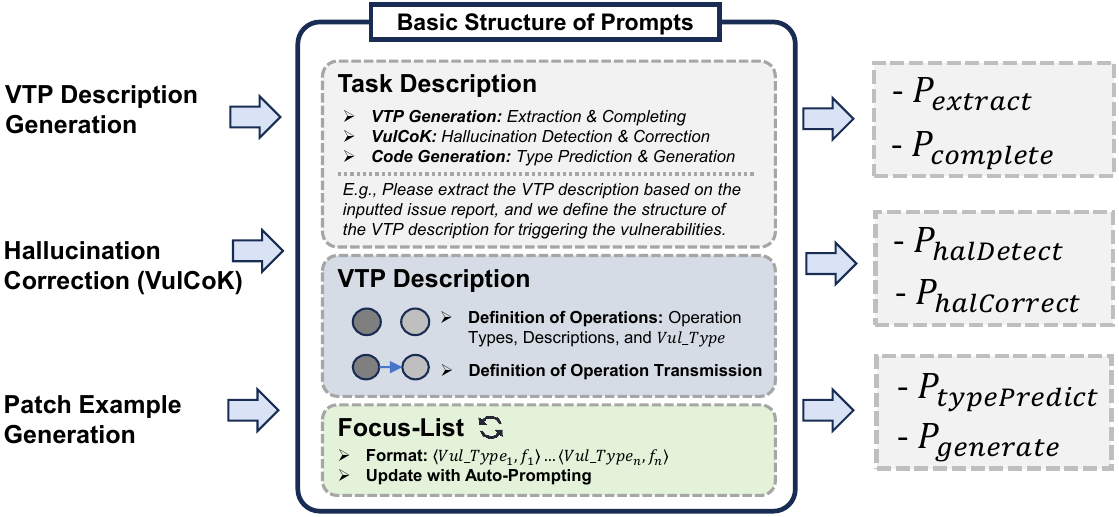}
\vspace{-0.7cm}
\caption{The prompt format $P_0$ in the {\tool}.}
\vspace{-0.2cm}
\label{fig:prompt_format}
\end{figure}

Based on the above definitions, we format the structure of the prompts $P_0$ in the {\tool}, as is shown in Figure \ref{fig:prompt_format}.
Each prompt consists of the following three components:
\ding{182} \textbf{Task Definition}, where we define the problem that LLM should resolve;
\ding{183} \textbf{Details of VTP Description}, where we incorporate the definition of the VTP operation, i.e., operation type, descriptions, and vulnerability type $Vul\_Type$, as well as the transmissions between edges. 
The $Vul\_Type$ contains two parts of the types, i.e., the error and CWE types, where Common Weakness Enumeration (CWE)~\cite{CWE} is a more specific categorization that differentiates the types of vulnerabilities with their causes and behaviors; 
and \ding{184} \textbf{Focus-List}, where we provide some focuses that guide LLM to resolve the problem.
The focus list contains a set of pairs $\langle Vul\_Type_i, f_i\rangle$, where 
$f_i$ is the focus of $Vul\_Type$ when generating VTP descriptions.
We introduce the focus list due to the following two reasons:

\begin{enumerate}[leftmargin=*]
    \item Some historical IRs may contain guidance on how to generate patch examples. LLMs can refer to this information to guide the generation of VTP descriptions and reduce output biases.
    \item Some IRs may lack detailed information to introduce the triggering process of vulnerability, so LLMs cannot directly generate the VTP description based on the contents. By referring to the focus list, LLMs can utilize historical information to complete the missing information in these IRs.
\end{enumerate}

The $f_i$ will be updated when LLM analyzes the historical IRs with the auto-prompting, and we will discuss the auto-prompting in the following sections in detail. 
To extract the VTP description, we will design the prompt $P_{extract}\leftarrow P_0$ by replacing the task description in this prompt format $P_0$ with instructions for VTP extraction.









\subsubsection{VTP Description Completing}\label{sec:vtp_completion}

Since some VTP descriptions are not complete and miss some essential operations and transitions in the paths, {\tool} will first detect whether the extracted VTP description is complete.
Then, it will complete the missing operation nodes and transitions. 
{\tool} detects the \textit{operation-level} and \textit{transition-level} completeness, and adds the missing information in the nodes and edges in these levels.    

\begin{itemize}[leftmargin=*]
    \item \textbf{Operation-level Completeness:} {\tool} detects whether the VTP description misses some intermediate operation nodes $Op_{miss}$, or the existing operation node $Op_i$ misses some information, such as the description of insecure code and $Vul\_Type$ misses the error or CWE types. {\tool} will ask the LLMs to reason the missing information within the operation nodes, or generate some new intermediate nodes between the existing operation nodes to complement the missing flows.
    \item \textbf{Transition-level Completeness:} The transitions between the operations $Op_i\rightarrow Op_j$ are missed, so the logic flow is not complete to trigger the vulnerability, and {\tool} will add these transition edges to complement the VTP description.
\end{itemize}

The prompt for VTP description $P_{complete}$ completion utilizes the same format $P_0$. The only difference is that we add the definition of the previous completeness. The prompt will also contain the different focus $f_i$ for different vulnerability types.

\subsubsection{Auto-Prompting for Generating VTP Description}

Since some of the LLMs cannot be directly fine-tuned, such as ChatGPT, the researchers have utilized the text generation ability of LLM to design the specific prompt for each input, i.e., Auto-Prompting~\cite{DBLP:conf/emnlp/ShinRLWS20}. 
We design a meta-framework for auto-prompting, as is shown in Algorithm \ref{alg:auto_prompting}. It utilizes the score function $F_s$ to calculate the differences between predicted and ground-truth labels in the labeled dataset and updates the $f_i$ in the prompt.
The line \ref{algo_line:update_prompt} indicates that the auto-prompting updates the prompt's focus $f_i$ by \textit{inserting}, \textit{deleting}, and \textit{modifying} elements in the original prompts. 
The auto-prompting process controls the prompt updating with simple prompts, such as "\textit{Please update the prompt by inserting|deleting|modifying the [item] to the prompt's focus $f$}", where \textit{[item]} is the sample used to optimize the prompt. 
The samples come from the historical IRs that can track the ground-truth insecure code and patch.
The line \ref{algo_line:vul_initialize}$\sim$\ref{algo_line:vul_final} indicate that {\tool} utilizes a score function $F_s$ to analyze the similarity between LLM outputs and ground-truth (\textit{lower the score, higher the similarity}). We select the most appropriate prompt $P_T$ based on the score differences among these three updated prompts.

\begin{algorithm}[t]
\small
	\caption{Process of Type-based Auto-Prompting.} 
 \label{alg:auto_prompting}
	\KwIn{The original prompt $P_0$, the dataset with labeled insecure codes and patches $LabeledDataset$, the predicted type $Vul\_Type$, and the score function for specific task $F_{s}$.} 
	\KwOut{The generated prompt $P_T$.}
    \SetKwProg{Fn}{Function}{}{end}
    $P_T\leftarrow P_0$, $Upd\_Prompts\leftarrow\{INSERT|DELETE|MODIFY\}$\;\label{algo_line:update_prompt}
    \Fn{Auto-Prompting($P_0$, $F_{s}$):}{
         $f=Focus[Vul\_Type]$\;\label{algo_line:vul_initialize}
        \For {$item\in LabeledDataset$}{
        $s_T=F_s(item.truth, item.pred,P_T)$\;
        $\{P_{insert}.f, P_{delete}.f, P_{modify}.f\}=LLM(P_T.f, Upd\_Prompts)$\;
        $P_{T}\leftarrow\{P_{insert}|P_{delete}|P_{update}\}$\;
        $P_T=\arg\max_{P_{T}}(s_T-F_s(item.truth, item.pred,P_{T}))$\;\label{algo_line:vul_final}
    }
    }
    return $P_T$\;
\end{algorithm}

The score function of auto-prompting the VTP description extractor and completer is calculated by analyzing the \textit{matching and masking scores}, which can be formulated as follows:
\begin{equation}
\resizebox{.91\linewidth}{!}{$
    \displaystyle
\begin{split}
    &F_s(VTP\_Code^{-}|VTP\_[M], VTP\_IR|VTP\_[M]', P_{extract}|P_{complete})=\\
    &\underbrace{sim(VTP\_Code^{-},VTP\_IR)}_{score_{match}}+\underbrace{sim(VTP\_[M],VTP\_[M]')}_{score_{mask}}
\end{split}
$}
\end{equation}
where $F_s$ is the score function that calculates the sum of two scores, i.e., $score_{match}$ and $score_{mask}$. The first score analyzes whether LLM can accurately generate the VTP descriptions that reflect the triggering process of vulnerabilities, and $socre_{mask}$ analyzes whether LLM can complement the incomplete IRs. 

For $score_{match}$, The $VTP\_IR$ is the generated IRs with original prompt $P_{extract}$, and $VTP\_Code^{-}$ is the ground-truth triggering path from the insecure code. We utilize the edit distance, i.e., Levenshtein Distance~\cite{DBLP:conf/ntms/AfzalGLB18}, as the similarity, which is useful to measure the similarity between two texts.
For $score_{mask}$, we randomly select some nodes in the extracted VTP, then reflect them to the original IR and mask these chosen texts $VTP\_[M]$. We utilize the LLM to predict the masked text to $VTP\_[M]'$ and calculate the edit distances between them.
We utilize these scores to measure the performance of LLM on generating VTP descriptions and update the prompts with the meta-framework.

\subsection{Correcting Hallucinatory VTP Description}

\begin{figure}[t]
\centering
\includegraphics[width=\columnwidth]{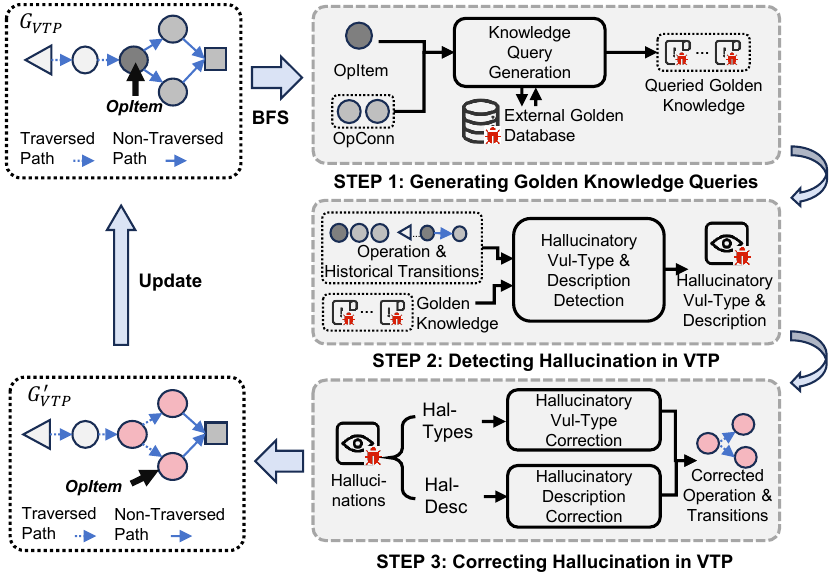}
\vspace{-0.7cm}
\caption{The logic flow of VulCoK.}
\label{fig:vulcok_flow}
\vspace{-0.5cm}
\end{figure}

In Huang et al's survey~\cite{DBLP:journals/corr/abs-2311-05232}, they indicate that the pre-trained data, training, and decoding strategies of LLMs have flaws that result in content that is inconsistent with real-world facts, which is called LLM hallucinations.
To address the hallucinations, Li et al.~\cite{DBLP:journals/corr/abs-2305-13269} proposed the CoK, which utilizes external golden knowledge for the hallucination correction.
Inspired by this work, we propose the VulCoK, as is shown in Figure \ref{fig:vulcok_flow}, which can correct the hallucinations in VTP operation nodes and transition edges.

\begin{table}[t]
\caption{The golden external dataset $\mathcal{D}$ of VulCoK.}
\vspace{-0.4cm}
\resizebox{\columnwidth}{!}{
\begin{tabular}{lccl}
\toprule
\textbf{Id} & \textbf{Golden Dataset}         & \textbf{Last Updated} & \multicolumn{1}{c}{\textbf{Link}}                            \\
\midrule
1 & SARD            & 2024         & \href{https://samate.nist.gov/SARD/}{https://samate.nist.gov/SARD/}                       \\
2 & OWASP           & 2024         & \href{https://owasp.org/www-project-benchmark}{https://owasp.org/www-project-benchmark}             \\
3 & Debian & 2024         & \href{https://bit.ly/3bX30ai}{https://bit.ly/3bX30ai}                              \\
4 & VDISC           & 2024         & \href{https://osf.io/d45bw/}{https://osf.io/d45bw/}     \\
\bottomrule
\end{tabular}}
\vspace{-0.5cm}
\label{tab:va_dataset}
\end{table}

\subsubsection{Hallucinatory VTP Detection}

The detection process of VTP description contains two parts. i.e., Vul-Type Hallucination Detection and Description Hallucination Detection, which detects the hallucinations in vulnerability types and descriptions of VTP nodes.
We first introduce the external golden databases Table \ref{tab:va_dataset}, which is selected based on the update time, the usage of databases in industry and research, and the number of vulnerabilities.
We detect the hallucinations in the VTP description with the Breadth-First Search (BFS)~\cite{DBLP:journals/ai/SiklossyRM73}, which searches for the current operation item $OpItem$ and its connected operations $\{OpConn|OpItem\rightarrow OpConn\}$. 
The LLM first generates the queries for retrieving the golden items in the dataset $\mathcal{D}$.
Similar to Section \ref{sec:vtp_completion}, we also utilize the operation and the historical transitions to analyze whether they contain the hallucination.  
We utilize the prompt $P_{halDetect}$ with the prompt format $P_0$ to detect the hallucination, which contains the definition of hallucinations, as well as the focus list of CWE and error types.

\subsubsection{Hallucinatory VTP Correction}
For the VTP descriptions that contain the hallucinations, i.e., the hallucinations from type and description in VTP. 
First, suppose the CWE and error types are incorrect and contain hallucinations. 
In that case, the VulCoK needs to correct the hallucinatory types in the node $OpItem$ and $OpConn$.
If the types are correct, the LLM needs to correct the hallucinations of the VTP description and $OpNext$ and re-generate the transitions for the new VTP operations.
After these corrections, the original $\mathcal{G}_{VTP}$ will be updated to $\mathcal{G}^{'}_{VTP}$, and the current item $OpItem$ will move to its connected items $OpConn$.
The prompt $P_{halCorrect}$ utilizes the same format $P_0$ for correcting the hallucinations. It directly asks LLM to correct the VTP's vulnerability types and descriptions based on the retrieved golden knowledge, and it will also incorporate the focus list of different vulnerability types.

\subsubsection{Auto-Prompting for VulCoK}

Since the retrieved knowledge in the selected golden dataset incorporated the insecure code, the score function of auto-prompting the VulCoK is calculated by analyzing the similarity between the tracked insecure code and the golden knowledge's insecure code.
\begin{equation}
\begin{split}
    &F_s(CodeGold^{-},Code^{-},P_{halDetect}|P_{halCorrect}) =\\
    &\sum_{ColdGold^{-}}sim(CodeGold^{-},Code^{-})
\end{split}
\end{equation}
where the $CodeGold^{-}$ indicates the insecure code in the golden knowledge, and the $sim$ is the edit distance. We sum all the distances in the retrieved knowledge, then we feed the $F_s$ into Algorithm \ref{alg:auto_prompting} and update the prompt with this meta-framework.


\subsection{Generating Insecure Code \& Patch Example}\label{sec:joint_code_patch_generation}
In this section, {\tool} first predicts the patch types based on the corrected VTP description. Then, it jointly generates the insecure coding \& patch examples based on the patch types.

\subsubsection{VTP-based Patch Type Prediction}

Previously, Chow et al.~\cite{DBLP:conf/icse/ChowGP24} defines 12 patch types for fixing the normal bugs in OSS projects. 
They construct a mapping from the error types to the patch types, which reflect which types of patches are more frequently used for fixing certain bugs.
Inspired by this work, we also ask the LLMs to predict the patch types before the patch example generation.
First, since the manual investigation of vulnerable IRs with patches shows the types of patches for fixing the vulnerabilities are similar to fixing the normal bugs, we directly migrate the patch type defined by Chow et al. to our patch type prediction.
Then, we ask the LLM to predict the patch type $Patch\_Type$ for the VTP description $\mathcal{G}_{VTP}$ and record the co-occurrence between the vulnerability type and patch type in the current predicted IRs $freq=\#(Vul\_Type,Patch\_Type)/\#Predicted\_IR$.
The prompt $P_{typePredict}$ incorporates the definition of patch types and the frequency, as well as the type and focus list in CWE and error types.

\subsubsection{Joint Insecure Code \& Patch Example Generation}

After the prediction of patch types, we generate the patch example based on the $Patch\_Type$ and the original VTP description $\mathcal{G}_{VTP}$. 
Since some nodes are operations in the imported third-party libraries, we first ask the LLMs to select the nodes and edges that reflect the developer's insecure coding process. 
Then, we utilize the selected nodes and edges to jointly generate the pairs with insecure coding and patch examples.
The idea of joint generation comes from multitask learning~\cite{DBLP:conf/kbse/LiSYW20}, where the incorrect output elements are modified based on other output elements' results, thus improving the accuracy of patch generation. 
The prompt $P_{generate}$ asks LLMs to select nodes/edges and jointly generate the pairs, as well as incorporates the focus list of CWE and error types

\subsubsection{Auto-Prompting for Patch Example Generation}

The auto-prompting process for the patch example generation utilizes the edit similarity between generated code and ground-truth code in the historical IRs to optimize the prompts, and the score function is shown as follows:
\begin{equation}
\begin{split} &F_s(Code^{-}|Patch^{+},Code'|Patch',P_{typePredict}|P_{generate}) = \\
&sim(Code',Code^{-})+sim(Patch',Patch^{+})
\end{split} 
\end{equation}
where the $Code'$ and $Patch'$ are the generated insecure coding and patch examples. The $Code^{-}$ and $Patch^{+}$ are the ground-truth of insecure code and patch of the vulnerabilities.
We also feed the $F_s$ into Algorithm \ref{alg:auto_prompting} and update the prompt with the meta-framework.
\section{Experimental Design}

To evaluate the performance of {\tool}, we will investigate the
following three research questions (RQs)

\begin{itemize}[leftmargin=*]
    \item \textbf{RQ1: How does {\tool} perform on generating insecure code examples?}
    \item \textbf{RQ2: How does {\tool} perform on generating patch examples for fix the vulnerability?}
    \item \textbf{RQ3: How does {\tool} handle IRs when they lack detailed information?}
    \item \textbf{RQ4: How does each component contributes to the {\tool} on generating insecure coding and patch examples?}
\end{itemize}

\subsection{Dataset Preparation}\label{sec:dataset_preparation}

In this section, we first enrich the IRs from original \textbf{GHArchive}~\cite{GHArchive} with other two representative data sources, i.e., \textbf{D2A}~\cite{DBLP:conf/icse/ZhengPLBEYLMS21} and \textbf{PatchDB}~\cite{DBLP:conf/dsn/WangWF0J21};
{then, we denoise the D2A dataset to improve its quality;}
third, we preprocess the dataset with token replacement and split the dataset into IRs for auto-prompting and evaluation.

\noindent\textbf{STEP 1: Collecting the Dataset.}
We collect the dataset from three major sources following previous works~\cite{DBLP:conf/sigsoft/PanZC0BHLH22,DBLP:journals/corr/abs-2008-04176}. The first data source is \textbf{GHArchive}~\cite{GHArchive}, a comprehensive dataset that contains over 120K GitHub IRs since 2015.
The second data source is \textbf{D2A}~\cite{DBLP:conf/icse/ZhengPLBEYLMS21}, which is built from real-world vulnerability prediction scenarios and contains over 10K insecure code found from GitHub IRs with their vulnerability types.
The third data source is \textbf{PatchDB}~\cite{DBLP:conf/dsn/WangWF0J21}, which incorporates over 4K security patches in the GitHub repositories.
All the datasets have been widely used in multiple vulnerability identification tasks~\cite{DBLP:journals/pieee/LinWHZX20,DBLP:conf/icmla/RussellKHLHOEM18,DBLP:conf/ndss/LiZXO0WDZ18,DBLP:conf/csr2/OmarS23,DBLP:conf/raid/0001DACW23,DBLP:conf/sigsoft/PanZC0BHLH22}.
We collect the vulnerability information in these two data sources by searching commit messages and vulnerable IRs, then we remove items without the searched vulnerable IRs.

\noindent\revise{\textbf{STEP 2: Denoising the Dataset.} The D2A is automatically built by the commit message analyzer, and the authors report that D2A only has 53\% accuracy in extracting commits.
Therefore, we remove 67 noisy samples from D2A as follows: \ding{182} we obtain the commit messages, and vulnerable IRs by manually searching the repositories; 
\ding{183} we remove the noisy items that the commit messages and IRs explicitly indicate that they do not contain the vulnerabilities; 
and \ding{184} in the remaining code, we remove the noisy items by checking whether the disclosed vulnerabilities are depreciated in the CVE.}
To reduce the biases in the data-denoising process, 
we have invited three security practitioners with over 5-year experience to determine whether the dataset is correctly denoised. We ask them to independently check whether the removed noisy items are accurate.
The average Cohen’s Kappa~\cite{DBLP:journals/jss/PerezDMT20} value is over 0.9, which means they highly agree on the noisy data removal.

\begin{table}[t]
\caption{The size of the dataset with three sources.}
\vspace{-0.4cm}
\resizebox{\columnwidth}{!}{
\begin{tabular}{l|cccc}
\toprule
\multirow{2}{*}{\textbf{Data Sources}}   & \multirow{2}{*}{\textbf{\#Vul-IRs}} & \multirow{2}{*}{{\textbf{\#Auto-Prompt}}} & \multicolumn{2}{c}{\textbf{\#Evaluation}} \\
&&&\textbf{w/ Labels}&\textbf{w/o Labels}\\
\midrule
\textbf{GHArchive}      & 4,316                     & 1,072 &  268                   & 2,976                     \\
\textbf{D2A}            & 662   &           530          & 132                        & -                         \\
\textbf{PatchDB}        & 487    &            390        & 97                        & -                         \\
\hline
\textit{\textbf{Total}} & \textit{5,465}   &        \textit{1,992}            & \textit{497}                       & \textit{2,976}\\
\bottomrule
\end{tabular}}
\vspace{-0.6cm}
\label{tab:dataset_size}
\end{table}



\noindent\textbf{STEP 3: Preprocessing the Dataset.} 
The GitHub IRs collected from the web pages are in XML format, 
and we need to preprocess the IRs by ragging screenshots and code snippets. 
We preprocess the IRs with the following procedures: 
\ding{182} 
{we first utilize the Tencent OCR to transit the screenshots (wrapped by XML tag \texttt{$<$a href=``.jpg|.png''$>$}) to the text~\cite{tecentOCR}, then use {{[SCR]}} to tag the screenshots, and {{[CODE]}} to tag the code snippets (wrapped by XML tags \texttt{$<$code$>$}, \texttt{$<$/code$>$}). The content will be {{[CODE]} \{content of code snippet\}} after tagging;
\ding{183} we merge similar code snippets and page screenshots, which may have few differences and describe similar vulnerability information;
and \ding{184} following the previous works~\cite{DBLP:conf/kbse/ShiJYCZMJW21}, we remove other XML tags and retain the plain text inside, then we correct typos with Spacy~\cite{spacy.io}.
We utilize 80\% of the IRs with code commits for auto-prompting the LLMs, and the rest of the IRs for evaluation. In consequence, the evaluation dataset contains IRs with/without code labels.}

{Table \ref{tab:dataset_size} shows the number of vulnerable IRs for auto-prompting and evaluation.
In total, we have obtained 5,465 vulnerable IRs from these three sources, where 1,992 for auto-prompting and 3,473 for evaluating the {\tool}.
Among the evaluation IRs, 2,976 IRs do not contain the tracked insecure code \& patches (\textbf{w/o Labels}).}

\subsection{Experimental Baselines}

%


\textbf{Non-LLM Baselines for Code Generation.}
\textbf{CodeBert}~\cite{DBLP:conf/emnlp/FengGTDFGS0LJZ20} is a large code model pre-trained on millions of code snippets with the BERT model. We fine-tune the CodeBert on the dataset for auto-prompting.
\textbf{Codeium}~\cite{codeium} is a low-cost AI-driven approach for code completion and searching.
We utilize these baselines to generate insecure code examples from the description of IR.
Compared with other baselines, they achieve SOTA performances in our task.

\noindent\textbf{Non-LLM Baselines for APR.}
The APR tools also utilize the fine-tuned \textbf{CodeBert} as the baseline.
\textbf{InCoder}~\cite{DBLP:conf/iclr/FriedAL0WSZYZL23} is designed for code infilling by adopting a causal masking objective.
We fine-tune these two models on the $\langle\textit{Insecure Code, Patch}\rangle$ pairs of evaluation dataset for auto-prompting.
To keep these APR baselines consistent with {\tool}, these two models generate the patches based on the generated insecure code example of {\tool}.
Compared with other baselines, they achieve SOTA performances in our task.


\noindent\textbf{Baselines with Generative LLMs.}
Recently, researchers have utilized the LLMs with prompt learning to automatically generate code and repair the bugs. 
We choose the three common LLM baselines in our tasks, which can achieve the SOTA performances.
\textbf{CodeT5}~\cite{DBLP:conf/emnlp/WangLGB0H23} is pre-trained on T5, which is an encoder-decoder model that takes into account token type information in the code.
\textbf{Codex (GPT-3)}~\cite{DBLP:conf/emnlp/YooPKLP21}
and \textbf{ChatGPT (i.e., GPT-3.5)}~\cite{LLMBackground} are two novel LLMs proposed by OpenAI, which use over 100B of parameters and are trained on over 10TB samples with multiple training strategies (few-shot, zero-shot, etc.).
We choose the stable and well-maintained versions: \textit{t5-base}~\cite{t5-version}, \textit{text-davinci-003}~\cite{gpt3_version}, and \textit{gpt-3.5-turbo}~\cite{gpt35_version}, and use the \textit{\textbf{same prompt in Section \ref{sec:joint_code_patch_generation}}}.
Except for the ChatGPT, all the baselines are \textit{\textbf{fine-tuned}} on our dataset, then generate code examples and predict types.

\subsection{Metrics and Experimental Settings}

\textbf{Metrics.}
The first metric is the \textbf{MatchLine}, which is a strict metric that measures the proportion of total matched statements with the ground-truth code.
The second is the \textbf{MatchTrig} and \textbf{MatchFix}, which measure the matching rate of statements that may contain insecure code (annotated with "-") and patch (annotated with "+"). These two metrics indicate whether the generated code can trigger or fix the vulnerabilities.
We choose the $K=10$ as the default value to measure these matching rates.
\textbf{AccType} is utilized to measure the accuracy of type prediction, and it measures the average of both CWE and error types in insecure code examples.
We also use the \textbf{Triggering Rate} (\textbf{Trig@$K$}) and \textbf{Fixing Rate} (\textbf{Fix@$K$}) to measure the triggering and fixing rate of generated code examples:
\begin{equation}
\resizebox{.91\linewidth}{!}{$
    \displaystyle
    Trig@K=\frac{\#Trig\_Vul@K}{\#Total\_Vul@K}, 
    Fix@K=\frac{\#(Trig\_Vul@K\cap Fix\_Vul@K)}{\#Total\_Vul@K}
$}
\end{equation}
where “\#” is the symbol of the number calculation of evaluation samples, and Fix@$K=1$ if both the vulnerability triggering and fixing are satisfied in the Top-$K$ generated pairs. 
We choose $K=1, 5, 10$ for measuring the triggering and fixing rates.


\noindent\textbf{Parameter and Hardware Settings.}
{We split 80\% of IRs with code commits for auto-prompting, and the rest 20\% and the IRs w/o commits for evaluation.
We fine-tune all the baselines (except for ChatGPT) with $batch\_size=8$.}
All experiments are run on a PC with Windows 11 OS, NVIDIA GeForce RTX 2060.
\section{Result}

\subsection{Performances on Insecure Code Generation}\label{sec:codegenresult}

We introduce the {\tool} to improve the T5, GPT-3, and ChatGPT's performances, and the model names are LLMs+{\tool}.
In the evaluation dataset with code labels, we analyze the matching rate between generated insecure code and the ground-truth labels of insecure code.
In the evaluation dataset without code labels, we \ding{182} first utilize the open-sourced security testing tools, such as Zed~\cite{zed} and Wapiti~\cite{wapiti}, etc., to test whether the generated insecure code example will trigger the corresponding vulnerabilities, and \ding{183} manually test the insecure code if the automatic detectors cannot trigger the vulnerabilities.


\noindent\textbf{Comparison Results.}
Table \ref{tab:res_code_gen}
illustrates the results of {\tool} on generating insecure code examples.
Comparing the {\tool} with all the code generation and LLM baselines, we can see that, the ChatGPT+{\tool} can obtain the highest performances with 74.3\% (MatchLine), 81.0\% (MatchTrig), and 80.6\% (Trig@10), improving LLM baselines with +37.1\% (MatchLine), +25.9\% (MatchTrig), and +10.5\% (Trig@10) on average.
Moreover, {\tool} also improves LLM baseline's accuracy in predicting the vulnerability types with +11.2\% on average.

\begin{table}[t]
\caption{The performances of baseline comparison on generating insecure code examples from vulnerable IRs (\%).}
\vspace{-0.4cm}
\resizebox{\columnwidth}{!}{\begin{tabular}{c|c|c|m{1.7cm}<{\centering}|lll}
\toprule
\textbf{Exp}                          & \textbf{Category}                              & {\textbf{Model}} & \textbf{Version}       & \textbf{MatchLine}                                               & \textbf{MatchTrig}                                            & \textbf{AccType}   
\\
\midrule
                \multirow{9}{*}{\makecell[c]{w/\\Code\\ Labels}}             &    \multirow{2}{*}{\makecell[c]{Non-LLM \\ CodeGen}}                                 & CodeBert                  &      -         & 16.6                                                    & 35.9                                                    & -                                                     \\
                             &  & Codeium                   &   1.6.10            & 45.2                                                    & 60.7                                                    & -                                                        \\
                             \arrayrulecolor{lightgray} \cmidrule{2-7}
                             &                                     & CodeT5                    & t5-base       & 18.2                                                    & 38.7                                                    & 66.3                                                     \\
                             
                             & \multirow{-2}{*}{T5}                & \cellcolor{gray!25}+{\tool}             & \cellcolor{gray!25}t5-base       & \cellcolor{gray!25}57.3 (\textcolor{red}{$\uparrow$}39.1)                                                    & \cellcolor{gray!25}68.2  (\textcolor{red}{$\uparrow$}29.5)                                                  & \cellcolor{gray!25}78.6   (\textcolor{red}{$\uparrow$}12.3)                                                  \\
                             \cmidrule{2-7}
                             &                                     & Codex                   & davinci-003     & 27.5                                                    & 50.2                                                    & 62.9                                                     \\
                             
                             & \multirow{-2}{*}{GPT-3}             & \cellcolor{gray!25}+{\tool}               & \cellcolor{gray!25}davinci-003     & \cellcolor{gray!25}66.2 (\textcolor{red}{$\uparrow$}38.7)                                                   & \cellcolor{gray!25}79.5 (\textcolor{red}{$\uparrow$}29.3)                                                   & \cellcolor{gray!25}80.7 (\textcolor{red}{$\uparrow$}17.8)                                                      \\
                             \cmidrule{2-7}
                             &                                     & ChatGPT                   & turbo-3.5 & 40.9                                                    & 62.1                                                    & 80.5                                                     \\
  & \multirow{-2}{*}{GPT-3.5}           & \cellcolor{gray!25}+{\tool}               & \cellcolor{gray!25}turbo-3.5 & \cellcolor{gray!25}\textbf{74.3}  (\textcolor{red}{$\uparrow$}33.4)                                                  & \cellcolor{gray!25}\textbf{81.0}  (\textcolor{red}{$\uparrow$}18.9)                                                    & \cellcolor{gray!25}\textbf{83.9}  (\textcolor{red}{$\uparrow$}3.4)                                                   \\
\arrayrulecolor{black}\midrule
\textbf{Exp}                          & \textbf{Category}                              & {\textbf{Model}} & \textbf{Version}       & {{{ \textbf{Trig@1}}}} & {{{\textbf{Trig@5}}}} & {{{\textbf{Trig@10}}}} \\
\midrule
                    \multirow{9}{*}{\makecell[c]{w/o\\Code\\ Labels}}         &    \multirow{2}{*}{\makecell[c]{Non-LLM \\CodeGen}}                                     & CodeBert                  &    -           & 39.2                                                    & 50.4                                                    & 61.3                                                     \\
                             & & Codeium                   &  1.6.10           & 20.5                                                    & 44.6                                                    & 47.9                                                     \\
                             \arrayrulecolor{lightgray} \cmidrule{2-7}
                             &                                     & CodeT5                    & t5-base       & 37.2                                                    & 55.6                                                    & 63.2                                                     \\
                             & \multirow{-2}{*}{T5}                & \cellcolor{gray!25}+{\tool}               & \cellcolor{gray!25}t5-base       & \cellcolor{gray!25}64.2 (\textcolor{red}{$\uparrow$}27.0)                                                   & \cellcolor{gray!25}68.6  (\textcolor{red}{$\uparrow$}13.0)                                                     & \cellcolor{gray!25}73.2  (\textcolor{red}{$\uparrow$}10.0)                                                      \\
                             \cmidrule{2-7}
                             &                                     & Codex                   & davinci-003     & 58.5                                                    & 63.4                                                    & 65.2                                                       \\
                             & \multirow{-2}{*}{GPT-3}             & \cellcolor{gray!25}+{\tool}               & \cellcolor{gray!25}davinci-003     & \cellcolor{gray!25}72.6  (\textcolor{red}{$\uparrow$}14.1)                                                     & \cellcolor{gray!25}74.2  (\textcolor{red}{$\uparrow$}10.8)                                                    & \cellcolor{gray!25}77.5   (\textcolor{red}{$\uparrow$}12.3)                                                     \\
                             \cmidrule{2-7}
                             &                                     & ChatGPT                   & turbo-3.5 & 67.0                                                      & 68.5                                                    & 71.3                                                     \\
 & \multirow{-2}{*}{GPT-3.5}           & \cellcolor{gray!25}+{\tool}               & \cellcolor{gray!25}turbo-3.5 & \cellcolor{gray!25}\textbf{73.5}    (\textcolor{red}{$\uparrow$}6.5)                                                   & \cellcolor{gray!25}\textbf{76.9}   (\textcolor{red}{$\uparrow$}8.4)                                                    & \cellcolor{gray!25}\textbf{80.6}   (\textcolor{red}{$\uparrow$}9.3)     \\
\arrayrulecolor{black}\bottomrule
\end{tabular}}
\vspace{-0.3cm}
\label{tab:res_code_gen}
\end{table}

\begin{figure}[b]
\centering
\includegraphics[width=\columnwidth]{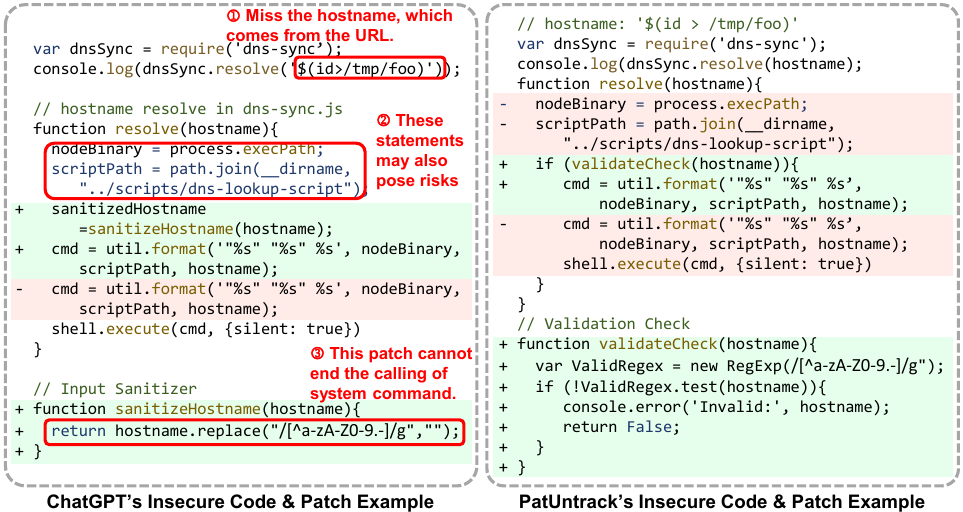}
\vspace{-0.4cm}
\caption{The case study of ChatGPT+{\tool} and ChatGPT on generating insecure code \& patch example.}
\label{fig:case_study}
\vspace{-0.3cm}
\end{figure}


\noindent\textbf{Case Study.}
We conduct the case study on the vulnerable IR in Figure \ref{fig:motivation_example} to qualitatively evaluate the {\tool}.
Figure \ref{fig:case_study} shows the generated insecure code example of ChatGPT and +{\tool}.
Referring to the ground-truth insecure code in the repository, we can see that {\tool} can accurately indicate that the \texttt{hostname} comes from the external URL and find all the statements that may contain CWE-78 vulnerabilities.
ChatGPT has made errors in these two points and generated inaccurate code examples.
These results illustrate that our approach is more accurate than baselines on generating insecure code examples.

\noindent\textbf{Advantages of {\tool}.}
We believe that the benefits of {\tool} come from three aspects.
\ding{182} First, {\tool} can obtain the description of how to trigger a vulnerability, and the VTP extractor and VTP completer improve the completeness of generated VTP, which can facilitate the LLMs to generate code that reflects the vulnerabilities. 
\ding{183} Second, the hallucination correction can reduce the VTP descriptions that do not reflect real-world vulnerabilities.
\ding{184} Third, the CWE and error type in the VTP description help LLM to accurately analyze the type of vulnerability.

\begin{center}
\small
\begin{tcolorbox}[colback=gray!10,
                  colframe=black,
                  width=\columnwidth,
                  arc=1mm, auto outer arc,
                  boxrule=0.5pt,
                  left=3pt,
                  right=3pt,
                  top=3pt,
                  bottom=3pt
                 ]
\textit{\textbf{Answering RQ1}: ChatGPT+{\tool} achieves the highest performances with 74.3\% (MatchLine), 81.0\% (MatchTrig), and 80.6\% (Trig@10). Moreover, {\tool} improves LLM baselines with +37.1\% (MatchLine) and +25.9\% (MatchTrig), and +10.5\% (Trig@10) on average.}
\end{tcolorbox}
\end{center}

\subsection{Performances on Patch Example Generation}
The experiment settings on the evaluation dataset with/without code labels are the same as Section \ref{sec:codegenresult}.
To keep Non-LLM APR baselines consistent with {\tool}, they generate the patches based on the generated insecure code example of {\tool}.

\begin{table}[t]
\caption{The performances of baseline comparison on generating patch examples from vulnerable IRs (\%).}
\vspace{-0.4cm}
\resizebox{\columnwidth}{!}{\begin{tabular}{c|c|c|c|lll}
\toprule
   \textbf{Exp} & \textbf{Category}                   & {\textbf{Model}} & \textbf{Version}       & \textbf{MatchLine} & \textbf{MatchFix} & \textbf{AccType} \\
\midrule
               \multirow{9}{*}{\makecell[c]{w/ \\ Code \\ Labels}}                        & \multirow{2}{*}{\makecell[c]{Non-LLM\\APR}} & {CodeBert}                         & -              & 41.9      & 56.5     & -     \\
                                       &  & {InCoder}                          & InCoder-6.8B  & 46.5      & 67.3     & -        \\
                                      \arrayrulecolor{lightgray} \cmidrule{2-7}
                                       & \multirow{2}{*}{T5}      & CodeT5                    & t5-base       & 40.7      & 59.7     & 68.9     \\
                                       &                          & \cellcolor{gray!25}+{\tool}             & \cellcolor{gray!25}t5-base       & \cellcolor{gray!25}51.2  (\textcolor{red}{$\uparrow$}10.5)    & \cellcolor{gray!25}74.0  (\textcolor{red}{$\uparrow$}14.3)     & \cellcolor{gray!25}77.2 (\textcolor{red}{$\uparrow$}8.3)    \\
                                       \cmidrule{2-7}
                                       & \multirow{2}{*}{GPT-3}    & Codex                   & davinci-003     & 53.0        & 62.2     & 76.2     \\
                                       &                          & \cellcolor{gray!25}+{\tool}          & \cellcolor{gray!25}davinci-003     & \cellcolor{gray!25}59.4  (\textcolor{red}{$\uparrow$}6.4)    & \cellcolor{gray!25}80.5 (\textcolor{red}{$\uparrow$}18.3)    & \cellcolor{gray!25}84.9 (\textcolor{red}{$\uparrow$}8.7)    \\
                                       \cmidrule{2-7}
                                       & \multirow{2}{*}{GPT-3.5} & ChatGPT                   & turbo-3.5 & 56.2      & 63.2     & 83.4     \\
                                       &                          & \cellcolor{gray!25}+{\tool}        & \cellcolor{gray!25}turbo-3.5 & \cellcolor{gray!25}\textbf{65.5}  (\textcolor{red}{$\uparrow$}9.3)    & \cellcolor{gray!25}\textbf{83.7} (\textcolor{red}{$\uparrow$}20.5)    & \cellcolor{gray!25}\textbf{87.2}  (\textcolor{red}{$\uparrow$}3.8)   \\
                                    \arrayrulecolor{black}   \midrule
\textbf{Exp} & \textbf{Category}                   & {\textbf{Model}} & \textbf{Version}       & \textbf{Fix@1}     & \textbf{Fix@5}    & \textbf{Fix@10}   \\
 \midrule
          \multirow{9}{*}{\makecell[c]{w/o \\ Code \\ Labels}}                           & \multirow{2}{*}{\makecell[c]{Non-LLM\\APR}}  & {CodeBert}                         &  -             & 38.4      & 51.5     & 54.2     \\
                                 &      & {InCoder}                          & InCoder-6.8B  & 56.6      & 60.7     & 62.1     \\
                                \arrayrulecolor{lightgray} \cmidrule{2-7}
                                       & \multirow{2}{*}{T5}      & CodeT5                    & t5-base       & 40.5      & 55.2     & 56.4     \\
                                       &                          & \cellcolor{gray!25}+{\tool}             & \cellcolor{gray!25}t5-base       & \cellcolor{gray!25}51.5 (\textcolor{red}{$\uparrow$}11.0)     & \cellcolor{gray!25}60.0  (\textcolor{red}{$\uparrow$}4.8)     & \cellcolor{gray!25}67.1  (\textcolor{red}{$\uparrow$}10.7)   \\
                                       \cmidrule{2-7}
                                       & \multirow{2}{*}{GPT-3}    & Codex                   & davinci-003     & 46.5      & 57.3     & 59.1       \\
                                       &                          & \cellcolor{gray!25}+{\tool}          & \cellcolor{gray!25}davinci-003     & \cellcolor{gray!25}66.2 (\textcolor{red}{$\uparrow$}19.7)     & \cellcolor{gray!25}72.3  (\textcolor{red}{$\uparrow$}15.0)   & \cellcolor{gray!25}75.9  (\textcolor{red}{$\uparrow$}16.8)   \\
                                       \cmidrule{2-7}
                                       & \multirow{2}{*}{GPT-3.5}                  & ChatGPT                   & turbo-3.5 & 50.3      & 61.7     & 62.3       \\
                                       &                          & \cellcolor{gray!25}+{\tool}        & \cellcolor{gray!25}turbo-3.5 & \cellcolor{gray!25}\textbf{69.7} (\textcolor{red}{$\uparrow$}19.4)     & \cellcolor{gray!25}\textbf{73.}2  (\textcolor{red}{$\uparrow$}11.5)   & \cellcolor{gray!25}\textbf{78.5} (\textcolor{red}{$\uparrow$}16.2)  \\
\arrayrulecolor{black}\bottomrule
\end{tabular}}
\vspace{-0.7cm}
\label{tab:res_patch_gen}
\end{table}

\noindent\textbf{Comparison Results.}
Table \ref{tab:res_patch_gen}
illustrates the comparison results of {\tool} on generating patch examples from IR textual description.
Comparing the {\tool} with all the code generation and LLM baselines, we can see that, the ChatGPT+{\tool} can obtain the highest performances with 65.5\% (MatchLine), 83.7\% (MatchFix), and 78.5\% (Fix@10), improving LLM baselines with +8.7\% (MatchLine), +17.7\% (MatchFix), and +14.6\% (Fix@10) on average.
Moreover, {\tool} also improves LLM's accuracy in predicting the patch types with +6.9\% on average.


\noindent\textbf{Case Study.}
Figure \ref{fig:case_study} also shows the results of the case study on patch example generation. 
We can see that the patch example generated by {\tool} can successfully utilize the input validation to fix the vulnerability. ChatGPT utilizes the \texttt{sanitizer} to replace the error strings with \texttt{null} but does not terminate the system command.
If the regular expression cannot match the whole input string, the attack may still succeed.
Therefore, the {\tool}'s patch example is more appropriate to fix the vulnerabilities.

\noindent\textbf{Advantages of {\tool}.}
In addition to the advantages in Section \ref{sec:codegenresult}, the benefits of {\tool} also come from the patch type prediction and the joint prediction. These methods can reduce the biases in generated patch examples, thus improving the accuracy and fixing rate of patch generation.

\begin{center}
\small
\begin{tcolorbox}[colback=gray!10,
                  colframe=black,
                  width=\columnwidth,
                  arc=1mm, auto outer arc,
                  boxrule=0.5pt,
                  left=3pt,
                  right=3pt,
                  top=3pt,
                  bottom=3pt
                 ]
\textit{\textbf{Answering RQ2}: ChatGPT+{\tool} achieves the highest performances with 65.5\% (MatchLine), 83.7\% (MatchFix), and 78.5\% (Fix@10).
Moreover, {\tool} improves LLM baselines with +8.7\% (MatchLine), +17.7\% (MatchFix), and +14.6\% (Fix@10) on average.}
\end{tcolorbox}
\end{center}

\subsection{Effect of IR's Detailed Information}

In Figure \ref{fig:prompt_format}, we introduce the focus list to guide the generation of patch examples when the IR lacks detailed information, and we also complete the missing nodes and edges in the VTP description.
In this experiment, we measure how the {\tool} depends on IR's detailed information.
Since there is no standard way to measure the richness of IR's information, we report the number of iterations (abbreviated as Iter.) in Section \ref{sec:vtp_generation},
where each iteration is a step with \textbf{VTP Extraction}$\leftrightarrows$\textbf{VTP Completing}.
This iteration indicates the difficulty of generating a completed VTP description. Intuitively, an IR with less information would require more iterations to generate a complete VTP description.

Based on the iterations, we split the evaluation IRs (\textbf{3,473 \#Evaluation} in Table \ref{tab:dataset_size}) into three intervals, i.e., \textbf{1$\leq$Iter.$<$4} (1,409/3,473 IRs), \textbf{4$\leq$Iter.$<$8} (1,161/3,473 IRs), and \textbf{Iter.$\geq$8} (903/3,473 IRs).
We can see that 26\% of them are minimally descriptive IRs (e.g., a single sentence to describe the vulnerability but lacks details), where the VTP generator takes $\geq$8 iterations to complete the missing nodes and edges in the generated VTP descriptions.

\begin{table}[b]
\caption{The performances on generating insecure code \& patch examples with different numbers of iterations(\%).}
\vspace{-0.4cm}
\resizebox{\columnwidth}{!}{\begin{tabular}{c|c|l|ll|ll}
\toprule
\multirow{2}{*}{\textbf{Iterations}}&\multirow{2}{*}{\textbf{Category}}&\multicolumn{1}{c|}{\multirow{2}{*}{\textbf{Models}}} &\multicolumn{2}{c|}{\textbf{Insecure Code Example}} & \multicolumn{2}{c}{\textbf{Patch Example}}\\
 & && \textbf{MatchTrig}    &  \textbf{Trig@10} & \textbf{MatchFix}     & \textbf{Fix@10} \\
\midrule
                         \multirow{6}{*}{1$\leq$Iter.$<$4}            &   & CodeT5      & 39.5         & 66.2                                 & 64.1         & 59.5                                \\
& \multirow{-2}{*}{\textbf{T5}}         & \cellcolor[HTML]{EFEFEF}+PatUntrack & \cellcolor[HTML]{EFEFEF}72.2 (\textcolor{red}{$\uparrow$}32.7) & \cellcolor[HTML]{EFEFEF}76.0 (\textcolor{red}{$\uparrow$}9.8)                          & \cellcolor[HTML]{EFEFEF}76.0 (\textcolor{red}{$\uparrow$}11.9) & \cellcolor[HTML]{EFEFEF}70.6 (\textcolor{red}{$\uparrow$}11.1)                        \\
\arrayrulecolor{lightgray} \cmidrule{2-7}
                   &          & Codex       & 56.5         & 66.4                                 & 64.9         & 63.2                                \\
& \multirow{-2}{*}{\textbf{GPT-3}}      & \cellcolor[HTML]{EFEFEF}+PatUntrack & \cellcolor[HTML]{EFEFEF}\textbf{83.1} (\textcolor{red}{$\uparrow$}26.6) & \cellcolor[HTML]{EFEFEF}78.6 (\textcolor{red}{$\uparrow$}12.2)                         & \cellcolor[HTML]{EFEFEF}82.4 (\textcolor{red}{$\uparrow$}17.5) & \cellcolor[HTML]{EFEFEF}77.2 (\textcolor{red}{$\uparrow$}14.0)                        \\
\cmidrule{2-7}
              &               & ChatGPT     & 70.5         & 82.0                                 & 66.4         & 67.4                                \\
& \multirow{-2}{*}{\textbf{GPT-3.5}}    & \cellcolor[HTML]{EFEFEF}+PatUntrack & \cellcolor[HTML]{EFEFEF}{82.5} (\textcolor{red}{$\uparrow$}12.0) & \cellcolor[HTML]{EFEFEF}\textbf{84.2} (\textcolor{red}{$\uparrow$}2.2)                          & \cellcolor[HTML]{EFEFEF}\textbf{85.0} (\textcolor{red}{$\uparrow$}18.6) & \cellcolor[HTML]{EFEFEF}\textbf{81.3} (\textcolor{red}{$\uparrow$}13.9)                        \\
\arrayrulecolor{black}\midrule
                        \multirow{6}{*}{4$\leq$Iter.$<$8}              &     & CodeT5      & 38.6         & 63.5                                 & 61.2         & 58.4                                \\
& \multirow{-2}{*}{\textbf{T5}}         & \cellcolor[HTML]{EFEFEF}+PatUntrack & \cellcolor[HTML]{EFEFEF}68.1 (\textcolor{red}{$\uparrow$}29.5) & \cellcolor[HTML]{EFEFEF}72.2 (\textcolor{red}{$\uparrow$}8.7)                          & \cellcolor[HTML]{EFEFEF}73.0 (\textcolor{red}{$\uparrow$}11.8) & \cellcolor[HTML]{EFEFEF}69.2 (\textcolor{red}{$\uparrow$}10.8)                        \\
\arrayrulecolor{lightgray} \cmidrule{2-7}

                &             & Codex       & 50.7         & 62.1                                 & 62.5         & 58.5                                \\
& \multirow{-2}{*}{\textbf{GPT-3}}      & \cellcolor[HTML]{EFEFEF}+PatUntrack & \cellcolor[HTML]{EFEFEF}\textbf{82.4} (\textcolor{red}{$\uparrow$}31.7) & \cellcolor[HTML]{EFEFEF}78.1 (\textcolor{red}{$\uparrow$}16.0)                         & \cellcolor[HTML]{EFEFEF}80.7 (\textcolor{red}{$\uparrow$}18.2) & \cellcolor[HTML]{EFEFEF}76.4 (\textcolor{red}{$\uparrow$}17.9)                        \\
 \cmidrule{2-7}
                &             & ChatGPT     & 61.5         & 78.4                                 & 65.5         & 61.5                                \\
& \multirow{-2}{*}{\textbf{GPT-3.5}}    & \cellcolor[HTML]{EFEFEF}+PatUntrack & \cellcolor[HTML]{EFEFEF}{81.9} (\textcolor{red}{$\uparrow$}20.4) & \cellcolor[HTML]{EFEFEF}\textbf{81.9} (\textcolor{red}{$\uparrow$}3.5)                          & \cellcolor[HTML]{EFEFEF}\textbf{85.6} (\textcolor{red}{$\uparrow$}20.1) & \cellcolor[HTML]{EFEFEF}\textbf{79.2} (\textcolor{red}{$\uparrow$}17.7)                        \\
\arrayrulecolor{black}\midrule
             \multirow{6}{*}{Iter.$\geq$8}              &                & CodeT5      & 11.0         & 31.9                                 & 23.5         & 28.7                                \\
& \multirow{-2}{*}{\textbf{T5}}         & \cellcolor[HTML]{EFEFEF}+PatUntrack & \cellcolor[HTML]{EFEFEF}67.4 (\textcolor{red}{$\uparrow$}56.4) & \cellcolor[HTML]{EFEFEF}71.5 (\textcolor{red}{$\uparrow$}39.6)                         & \cellcolor[HTML]{EFEFEF}73.2 (\textcolor{red}{$\uparrow$}49.7) & \cellcolor[HTML]{EFEFEF}65.4 (\textcolor{red}{$\uparrow$}36.7)                        \\
\arrayrulecolor{lightgray} \cmidrule{2-7}

                 &            & Codex       & 25.6         & 36.7                                 & 39.0         & 30.0                                \\
& \multirow{-2}{*}{\textbf{GPT-3}}      & \cellcolor[HTML]{EFEFEF}+PatUntrack & \cellcolor[HTML]{EFEFEF}76.2 (\textcolor{red}{$\uparrow$}50.6) & \cellcolor[HTML]{EFEFEF}76.9 (\textcolor{red}{$\uparrow$}40.2)                         & \cellcolor[HTML]{EFEFEF}76.4 (\textcolor{red}{$\uparrow$}37.4) & \cellcolor[HTML]{EFEFEF}73.6 (\textcolor{red}{$\uparrow$}43.6)                        \\
\cmidrule{2-7}
               &              & ChatGPT     & 43.7         & 65.4                                 & 54.2         & 49.5                                \\
& \multirow{-2}{*}{\textbf{GPT-3.5}}    & \cellcolor[HTML]{EFEFEF}+PatUntrack & \cellcolor[HTML]{EFEFEF}\textbf{80.1} (\textcolor{red}{$\uparrow$}36.4) & \cellcolor[HTML]{EFEFEF}\textbf{79.5} (\textcolor{red}{$\uparrow$}14.1)                         & \cellcolor[HTML]{EFEFEF}\textbf{80.2} (\textcolor{red}{$\uparrow$}26.0) & \cellcolor[HTML]{EFEFEF}\textbf{76.8} (\textcolor{red}{$\uparrow$}27.3)   \\
\arrayrulecolor{black}\bottomrule
\end{tabular}}
\end{table}

We compare the performances of LLMs and LLM+{\tool} on the metrics of insecure code example generation (i.e., MatchTrig and Trig@10), and patch example generation (i.e., MatchFix and Fix@10) within these three iteration intervals.
We can see that the performances of all techniques are decreased when the information in the IR is less detailed. 
However, {\tool} can accurately generate the insecure code \& patch examples, outperforming these original LLMs, with over +14.1\% (Trig@10) and +27.3\% (Fix@10) in Iter.$\geq$8. 
Moreover, {\tool} has fewer fluctuations than original LLMs among different iteration intervals.
From 1$\leq$Iter.$<$4 to Iter.$\geq$8, the performances of {\tool} decrease by less than $\pm$5.0\% (Trig@10) and $\pm$6.0\% (Fix@10).
These results illustrate the ability of {\tool} to handle IRs that lack detailed information.

\begin{center}
\small
\begin{tcolorbox}[colback=gray!10,
                  colframe=black,
                  width=\columnwidth,
                  arc=1mm, auto outer arc,
                  boxrule=0.5pt,
                  left=3pt,
                  right=3pt,
                  top=3pt,
                  bottom=3pt
                 ]
\textit{\textbf{Answering RQ3}: {\tool} can handle the IRs when they lack detailed information. It outperforms LLM baselines with over +14.1\% (Trig@10) and +27.3\% (Fix@10) when Iter.$\geq$8.
It also has the fewer fluctuations among iteration intervals with less than $\pm$5\% (Trig@10) and $\pm$6\% (Fix@10).}
\end{tcolorbox}
\end{center}

\subsection{Ablation Study}\label{sec:ablation}

In the ablation study, we conduct experiments on four types of variants, i.e., VTP generator, VulCoK, patch example generator, and auto-prompting. We compare the performances of {\tool} and variants on \textbf{MatchTrig} and \textbf{MatchFix}:
\begin{itemize}[leftmargin=*]
    \item \textbf{VTP Generator:} The variants are removing the Vul\_Type, VTP completer, and the whole VTP extractor.
    \item \textbf{VulCoK:} The variants are replacing VulCoK with CoK/ReAct, and removing the whole VulCoK.
    \item \textbf{Patch Example Generator:} The variants are removing the Vul\_Type, Joint Generation, and both variants.
    \item \textbf{Auto-Prompting:} The variants are replacing VulCoK with In-Context Learning (ICL)~\cite{DBLP:conf/emnlp/MinLHALHZ22}, which is a representative method that optimizes ChatGPT with relevant samples, and removing the Focus-List or the whole Auto-Prompting.
\end{itemize}

\begin{figure}[b]
\centering
\includegraphics[width=\columnwidth]{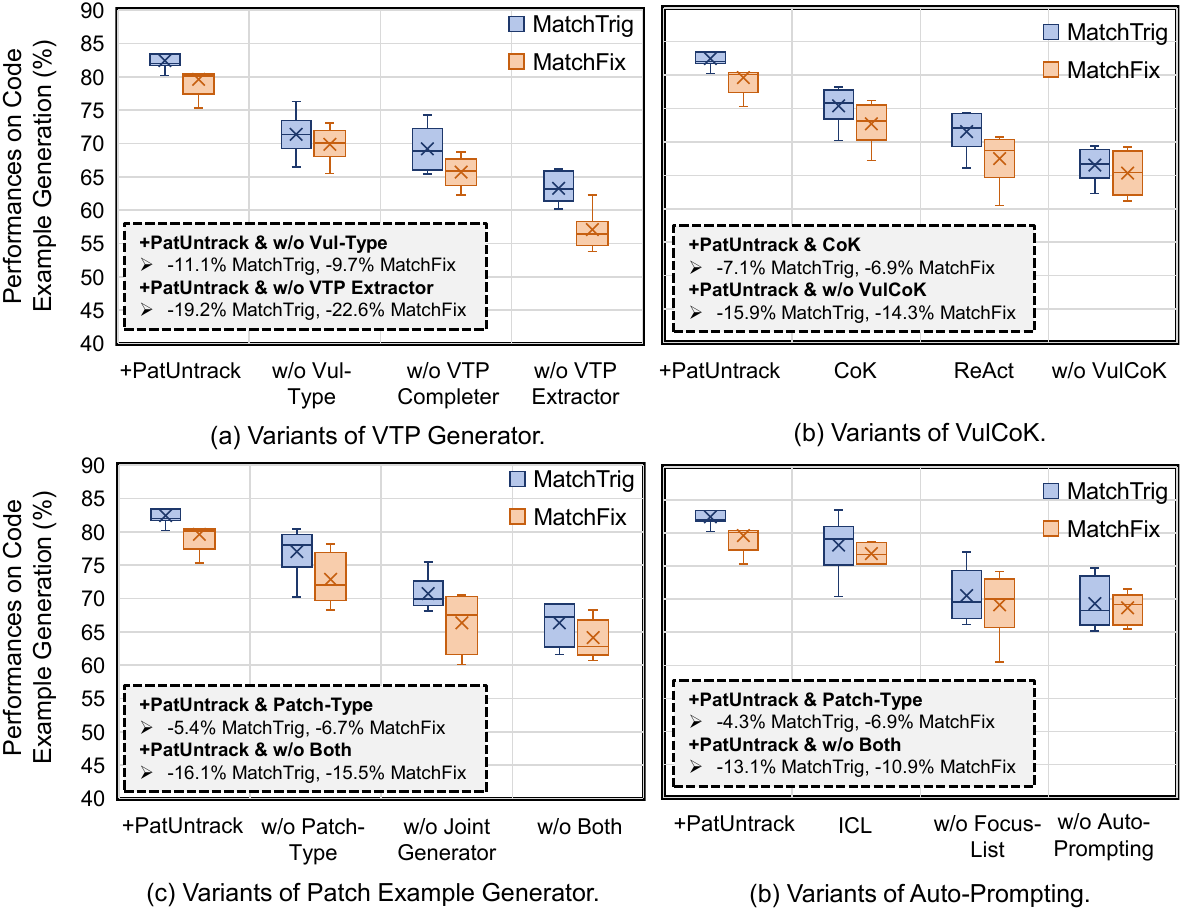}
\vspace{-0.4cm}
\caption{The results of the ablation study ({\tool} and variants utilize the ChatGPT as the basic model).}
\label{fig:ablation_study_with}
\vspace{-0.3cm}
\end{figure}

\noindent\textbf{Comparison on VTP Generator.} Figure \ref{fig:ablation_study_with} (a) shows comparison results of the VTP generator. 
We can see that removing vulnerable types leads to a moderate decrease of -11.1\% (MatchTrig) and -9.7\% (MatchFix); removing the whole VTP extractor leads to the largest decrease with -19.2\% (MatchTrig) and -22.6\% (MatchFix).

\noindent\textbf{Comparison on VulCoK.} 
Figure \ref{fig:ablation_study_with} (b) shows comparison results of VulCoK. 
We can see that replacing it with normal CoK leads to a moderate decrease of -7.1\% (MatchTrig) and -6.9\% (MatchFix); removing the whole VulCoK leads to the largest decrease with -15.9\% (MatchTrig) and -14.3\% (MatchFix).

\noindent\textbf{Comparison on Patch Example Generator.} 
Figure \ref{fig:ablation_study_with} (c) shows comparison results of the insecure code \& patch generator. 
We can see that removing patch types leads to a moderate decrease with -5.4\% (MatchTrig) and -6.7\% (MatchFix); removing both patch type and joint generation leads to the largest decrease with -16.1\% (MatchTrig) and -15.5\% (MatchFix).

\noindent\textbf{Comparison on Auto-Prompting} Figure \ref{fig:ablation_study_with} (d) shows the comparison results of auto-prompting.
We can see that replacing it with ICL leads to a moderate decrease of -4.3\% (MatchTrig) and -6.9\% (MatchFix); removing the auto-prompting leads to the largest decrease with -13.1\% (MatchTrig) and -10.9\% (MatchFix).

\begin{center}
\small
\begin{tcolorbox}[colback=gray!10,
                  colframe=black,
                  width=\columnwidth,
                  arc=1mm, auto outer arc,
                  boxrule=0.5pt,
                  left=3pt,
                  right=3pt,
                  top=3pt,
                  bottom=3pt
                 ]
\textit{\textbf{Answering RQ4}: {\tool} outperforms all the variants in ablation study. 
Removing the VTP extractor, VulCoK, joint patch generator, and auto-prompting leads to the largest decrease on MatchTrig and MatchFix; removing vulnerability/patch types and replacing components with CoK/ICL leads to a moderate decrease.}
\end{tcolorbox}
\end{center}

\section{Human Evaluation}

\begin{table}[b]
\caption{The number (\#) and the ratio of accepted generated insecure code \& patches for ChatGPT and {\tool}.}
\vspace{-0.4cm}
\resizebox{\columnwidth}{!}{
\begin{tabular}{l|rr|rrr}
\toprule
\textbf{CWE-Types} & \textbf{\#Total-Pairs} & \textbf{\#Response-Pairs} &  \textbf{\#Acc-{\tool}} & \textbf{\#Acc-ChatGPT} & \textbf{\#Acc-Both}  \\
\midrule
\textbf{CWE-79}             & 20             & 12 (60.0\%)             & 9 (45.0\%)             & 4 (20.0\%)                & 2 (10.0\%)           \\
\textbf{CWE-787}            & 17             & 8 (47.1\%)              & 4 (23.5\%)             & 4 (23.5\%)                & 3 (17.6\%)           \\
\textbf{CWE-78}             & 10             & 7 (70.0\%)              & 4 (40.0\%)             & 2 (20.0\%)                & 2 (20.0\%)           \\
\textbf{CWE-352}            & 8              & 3 (37.5\%)              & 3 (37.5\%)             & 1 (12.5\%)                & 1 (12.5\%)           \\
\textbf{CWE-287}            & 8              & 3 (37.5\%)              & 2 (25.0\%)             & 2 (25.0\%)                & 2 (25.0\%)           \\
\textbf{CWE-121}            & 7              & 2 (28.6\%)              & 2 (28.6\%)             & 1 (14.3\%)                & 1 (14.3\%)           \\
\textbf{CWE-119}            & 6              & 2 (33.3\%)              & 3 (50.0\%)             & 1 (16.7\%)                & 1 (16.7\%)           \\
\hline
\textbf{\textit{Total}}     & \textit{76}    & \textit{37 (48.7\%)}    & \textit{27 (35.5\%)}   & \textit{15 (19.7\%)}      & \textit{12 (15.8\%)}\\
\bottomrule
\end{tabular}
}
\vspace{-0.3cm}
\label{tab:acceptance}
\end{table}

To analyze the performances of {\tool} on generating patch examples for the vulnerable IRs without insecure code, we track the vulnerable IRs from GitHub that are \textbf{not included in the collected dataset} in Section \ref{sec:dataset_preparation}.
We track the IRs with the issue-tracking system~\cite{DBLP:conf/sigsoft/PanZC0BHLH22}
Then, we observe that 76 CVE-disclosed vulnerable IRs have not released the code commits, which may pose security risks to the public.
To analyze the contribution of {\tool}, we utilize the ChatGPT and ChatGPT+{\tool} to generate the patch examples and ask the authors the following questions:
\begin{itemize}[leftmargin=*]
    \item \textit{\textbf{Q1:} Does the type of patch example we provide match the vulnerabilities you have encountered? Please reply: \{Yes/No\}.}
    \item \textit{\textbf{Q2:} Will you accept the patch example to help you fix the vulnerabilities in your projects? Please reply: \{Accepted/Unaccepted\}.}
    \item \textit{\textbf{Q3:} Please use several sentences to describe the reason why the patch example can/cannot be accepted to fix the vulnerabilities.}
\end{itemize}
We contact the IR authors with \textit{Emails} or directly submit the comments on the IR pages. Then, we analyze the proportion of acceptance.
Table \ref{tab:acceptance} shows the number and ratio of accepted insecure code \& patch examples. 
We received 37 of 76 responses (48.7\%) from the IR authors, and 27 pairs (35.5\%) generated by ChatGPT+{\tool} can help authors fix the vulnerabilities, which is +15.8\% higher than ChatGPT.
Most insecure code \& patch example pairs (20) belong to CWE-79, and 9 pairs (45.0\%) are accepted.

Moreover, we ask the authors to manually inspect the generated results and rate their "\textit{accurateness}". This criterion measures the accuracy of the generated insecure code \& patch examples to the real commits in their projects. 
We ask the authors to rate 1-5 under the above criteria. For each accepted/unaccepted patch example, a score of 5 means that the generated code almost matches the real code in the repository, and a score of 1 means the generated code is completely different from the real code. A score of 3 is borderline, which means the generated code only needs a few modifications to become satisfactory code and reflect the vulnerabilities.

\begin{figure}[t]
\centering
\includegraphics[width=\columnwidth]{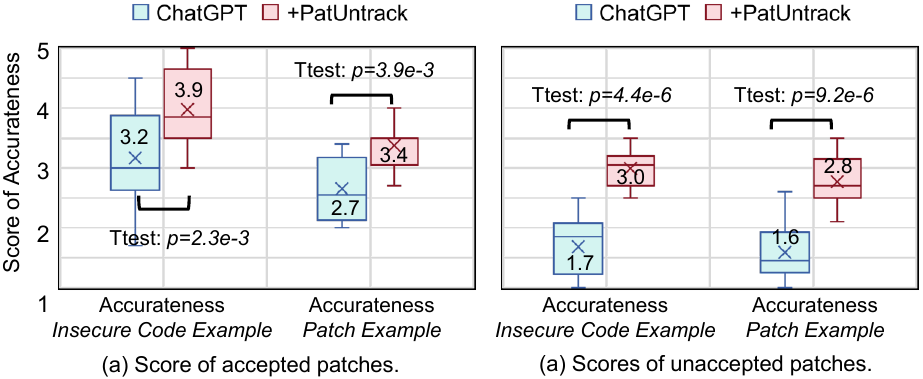}
\vspace{-0.6cm}
\caption{The score distribution of human evaluation ({\tool} utilizes the ChatGPT as the basic model).}
\label{fig:human_evaluation_score}
\vspace{-0.5cm}
\end{figure}

Figure \ref{fig:human_evaluation_score} shows the scores of accepted and unaccepted pairs for both ChatGPT and ChatGPT+{\tool}. 
For the accepted pairs, the scores of +{\tool} are significantly higher than ChatGPT\footnote{$p<0.05$ in the T-test shows the significant differences between two sets of data.}, and the improvement of average scores are both +0.7.
For the unaccepted pairs, +{\tool} also significantly outperforms the ChatGPT; the improvement scores are +1.3 and +1.2 on average.
The average scores of +{\tool} are nearly 3.0, which means the unaccepted code only needs a few corrections to be accepted.
These results illustrate that {\tool} can be practically utilized to help authors fix their vulnerabilities.

\section{Discussion}

\subsection{Effect of Joint Code Generation}\label{sec:paramaeters}
In Section \ref{sec:joint_code_patch_generation}, we utilize the joint code generation to generate the insecure code \& patch examples accurately.
To analyze the effect of insecure code examples on the patch example generation during the auto-prompting, we compare the performances of ChatGPT and ChatGPT+{\tool} on the change of \textbf{MatchTrig}$\rightarrow$\textbf{MatchFix} and \textbf{Trig@10}$\rightarrow$\textbf{Fix@10} on our evaluation dataset.
\begin{figure}[b]
\centering
\includegraphics[width=\columnwidth]{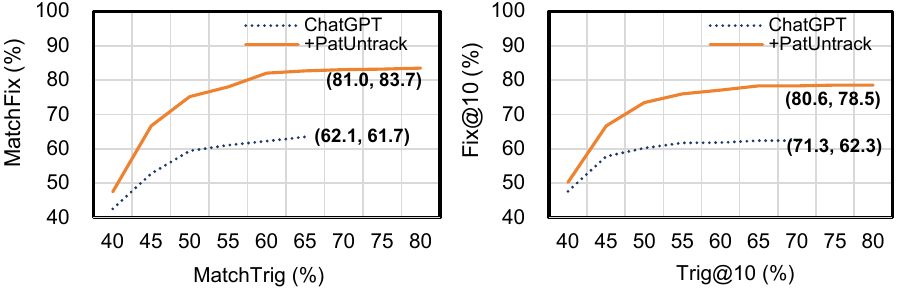}
\vspace{-0.6cm}
\caption{Insecure code's effect to patch example generation.}
\label{fig:parameter_change}
\vspace{-0.3cm}
\end{figure}

Figure \ref{fig:parameter_change} shows the generation results of insecure code \& patch examples in the auto-prompting.
We can see that, with the improvement of MatchTrig and Trig@10, both ChatGPT and {\tool}'s generation results on patch examples will improve, and the differences between these models will also increase.
When MatchTrig and Trig@10 reach 60\%$\sim$65\%, {\tool}'s MatchFix and Fix@10 reach the maximum value, which means patch examples are appropriate based on the insecure code examples.

\subsection{Unsuccessful Cases}
{Although {\tool} can generate appropriate insecure code \& patch examples, there are still 10\%$\sim$15\% cases, where \tool\ failed to generate correct patches. 
We manually inspected these unsuccessful cases and found that most of these cases come from incompletely generated VTPs.
Some IR authors omit too many details in the IR descriptions, so {\tool} cannot complete the VTP operations and transitions solely based on the IR descriptions.}
For example, the \textit{bedita/bedita/issues/755}~\cite{bad_case_url} (CVE-2015-9260) 
does not provide any description of the vulnerability except for links to external reports that are currently not analyzed by \tool, thus \tool\ cannot generate insecure code \& patch examples based on IR description.
{In the future, we plan to improve {\tool} by enhancing the input source for the VTP description.}

\subsection{{Threats to Validity}}\label{sec:threats}


\textbf{Internal Threats.}
The internal threat mainly comes from the approach. 
First, we only analyze the textual description in IR to build the VTP description, and we plan to utilize other sources, such as textual description in the third-party links to improve the {\tool}.
Second, the time costs of loops in the VTP generator and VulCoK are heavy, which may affect its practical usage.
To alleviate it, in addition to optimizing LLM with auto-prompting, we set an upper limit $\theta=10$ for each loop, which means the current step will exit if the loops exceed this limitation.
Third, we utilize the LLM to correct hallucinatory types and descriptions, which may have biases in the VulCoK.
In the future, we plan to analyze the success rate of hallucination correction to mitigate this threat.

\noindent\textbf{External Threat.}
The external threat may come from the dataset we use.
We only refer to the CVE-disclosed IRs to analyze whether these IRs contain the vulnerabilities.
However, some vulnerable IRs may not be incorporated by these two datasets, and some vulnerabilities may be disclosed by other security databases, such as CAPEC~\cite{CAPEC1}.
Another threat comes from the silent patches, since the authors may quietly submit patches for the vulnerabilities but not report them to the public.  
To alleviate it, we manually inspected 100 vulnerable IRs from the public and found that only 7 of them have these threats, so the impact of external threats is small.

\noindent\textbf{Constructive Threat.}
The constructive threat mainly comes from the metrics. All the chosen metrics, i.e., MatchLine, MatchTrig/MatchFix, and Trig\&$K$/Fix@$K$, may have biases for evaluating the generated results.
We review and discuss the settings of metrics with team members in the dataset preparation and experiments, thus alleviating this constructive threat.
\section{Related Works}

\textbf{Vulnerability Detection and Analysis from OSS Projects.}
Recently, researchers have proposed various approaches to detect and analyze the vulnerabilities in OSS projects.
Automatic vulnerability detection aims to determine whether there are malicious code in the projects~\cite{DBLP:journals/nca/ZhuLSZ23,liu2012software,malhotra2015systematic,ghaffarian2017software,ji2018coming,jie2016survey,DBLP:journals/pieee/LinWHZX20}.  
The researchers first proposed the statistic, dynamic, and hybrid techniques to detect the vulnerabilities with rules~\cite{DBLP:conf/sosp/EnglerCC01,DBLP:conf/snpd/LiangWWX16,DBLP:conf/sp/JangAB12}.
With the development of machine learning (ML) and Deep Learning (DL) approaches, researchers utilized these novel models to automatically build code features and improve the efficiency of vulnerability detection tools~\cite{DBLP:conf/ccs/YamaguchiWGR13,DBLP:journals/tse/ScandariatoWHJ14,DBLP:journals/tse/ShinMWO11,DBLP:conf/ndss/LiZXO0WDZ18,DBLP:conf/ndss/LiZXO0WDZ18,10.1145/3133956.3138840,DBLP:journals/corr/abs-2309-15324}.
In addition, researchers also analyze the vulnerabilities from various project artifacts (e.g., IRs, bug reports, etc.). 
Some researchers utilized text-mining methods to explore the security bug reports to identify the vulnerabilities~\cite{DBLP:conf/msr/GegickRX10,DBLP:conf/iecon/WijayasekaraMM14,DBLP:conf/hsi/WijayasekaraMWM12,DBLP:conf/kbse/WuZDYYCLJ20}, while other works analyze the negative impact of the vulnerabilities from the IRs~\cite{DBLP:journals/sqj/OyetoyanM21,DBLP:journals/tse/PetersTYN19,DBLP:journals/ese/ShuXCWM21,DBLP:conf/sigsoft/PanZC0BHLH22}.
The other researchers focus on the crowd-based security discussions, e.g., security posts in Stack Overflow, and discussion groups in Gitter/Slacks, to analyze the topics, attacks, and the corresponding mitigations~\cite{pletea2014security,meyers2019pragmatic,le2021large,yang2016security,zahedi2018empirical}.
Our work {\tool} is different from these previous works.
We build the gaps between the IR textual description and source code by generating insecure code \& patch examples for IRs that cannot track the insecure code, thus helping developers fix the vulnerabilities.

\noindent\textbf{Patch Generation for Vulnerable OSS Projects.}
APR methods are the typical methods for generating patches for fixing normal bugs or vulnerabilities.
The template-based APR methods leverage different bug-fixing templates, which are designed by human experts, to fix the specific types of bugs in the source code~\cite{DBLP:conf/issta/LiuK0B19,DBLP:conf/kbse/GhanbariZ19,DBLP:conf/issta/GhanbariBZ19}.
Recent researchers have proposed learning-based APR tools, which typically model program
repair as a Neural Machine Translation (NMT) problem~\cite{DBLP:conf/sigsoft/ZhuSXZY0Z21,DBLP:conf/icse/JiangL021,DBLP:conf/icse/YeMM22,DBLP:conf/issta/LutellierPPLW020}.
With the development of LLM, the researchers also analyze how to combine the LLMs to the APR tools to improve their patching ability~\cite{DBLP:conf/icse/XiaWZ23}, and reduce the time and financial cost of LLM~\cite{DBLP:journals/corr/abs-2304-00385}.
These works rely on the source code to fix the vulnerabilities, which cannot be applied on IRs without tracked insecure code.
On the contrary, our approach can generate patch examples based on IR textual description, which can timely help developers fix the vulnerabilities after the IR creation.
\section{Conclusion}
In this paper, we introduced {\tool} to generate patch examples from IRs without tracked insecure code.
It auto-prompts LLMs to make them applicable for analyzing the vulnerabilities described in IRs and generating appropriate patch examples.
Specifically, it first generates the completed VTP description from vulnerable IRs.
Then, it utilizes the VulCoK to correct the hallucinatory VTP description.
Finally, it generates Top-$K$ pairs of \textit{Insecure Code and Patch Example} based on the corrected VTP description.
Experiments conducted on 5,465 vulnerable IRs show that {\tool} can achieve the
highest performance and improve the traditional LLM baselines by +17.7\% (MatchFix) and +14.6\% (Fix@10) on average in patch example generation.
Furthermore, {\tool} has been applied to generating patch examples for 76 newly disclosed vulnerable IRs, and 27 out of 37 replies from the authors of these IRs confirmed the usefulness of the patch examples generated by {\tool}, indicating that they can benefit from these examples for patching the vulnerabilities.

In the future, we plan to enhance the {\tool} by introducing other third-party resources to generate the VTP descriptions, as well as analyzing whether VTP can improve the performance of traditional APR tools in patch generation.
\section*{Acknowledgments}
This work is supported by 
the National Natural Science Foundation of China Grant No. 62272445, 62232016, 62072442, the Youth Innovation Promotion Association Chinese Academy of Sciences,
and the University of Queensland NSRSG Grant NS-2201.


\bibliographystyle{ACM-Reference-Format}
\bibliography{ref}


\end{document}